%% file: cd17-arXiv-v2.tex
\documentclass[a4paper]{article} 
\usepackage[english]{babel}
\usepackage[utf8]{inputenc}
\usepackage[babel]{csquotes}
\usepackage[pass]{geometry}
\usepackage{amsmath,amssymb}
\usepackage{mathtools}
\usepackage{enumitem}
\usepackage[colorlinks, citecolor=blue, linkcolor=blue, linktoc=page,urlcolor=blue]{hyperref}
\usepackage{cool}
\usepackage{cite}
\usepackage{enumerate}
\usepackage{graphicx}
\usepackage{array}
\usepackage{subfigure}
\usepackage{verbatim}
\usepackage{fancyhdr}
\usepackage{lastpage}
\usepackage{epstopdf}
\usepackage{simplewick}
\usepackage{cancel}
\usepackage{epigraph}
\usepackage{tikz}
\usepackage{feynmf}
\usepackage{microtype}
\usepackage[toc,page]{appendix}
\DeclareMathOperator{\arcsinh}{arcsinh}

\usetikzlibrary{decorations.markings}
\usetikzlibrary{shapes.misc}
\tikzset{->-/.style={decoration={markings,mark=at position #1 with {\arrow{>}}},postaction={decorate}},->-/.default={0.5}}
\tikzset{-<-/.style={decoration={markings,mark=at position #1 with {\arrow{<}}},postaction={decorate}},-<-/.default={0.5}}
\tikzset{cross/.style={cross out, draw=black, minimum size=2*(#1-\pgflinewidth), inner sep=0pt, outer sep=0pt},cross/.default={1pt}}

\date{} 

\def\keywords#1{\begin{center}{\bf Keywords}\\{#1}\end{center}} %

\begin{document}
\title{Time-dependent Features in the Primordial Spectrum}
\author{Laura Covi$^{(a)}$ and Simone Dresti$^{(b)}$\\ 
       Institute for Theoretical Physics \\
       Faculty of Physics \\
       Georg-August University G\"ottingen \\ \\
       $^{(a)}$ \tt{laura.covi@theorie.physik.uni-goettingen.de } \\
        $^{(b)}$ \tt{simone.dresti@theorie.physik.uni-goettingen.de} 
       }
\maketitle
\thispagestyle{empty}

\begin{abstract}
In a Quantum Field Theory with a time-dependent background, time-translational symmetry is broken. 
We therefore expect time-dependent loop corrections to cosmological observables after renormalization
for an interacting field, with the consequent physical implications. 
In this paper we compute and discuss such radiative corrections to the primordial spectrum within simple 
models, both for massless and massive virtual fields, and we disentangle the time-dependence caused by 
the background and by the initial state after renormalization. 
For the investigated models the departure from near-scale-invariance is very small and there is full compatibility 
with the current Planck data constraints. Future CMB measurements may improve the current constraints on 
feature-full primordial spectra and possibly observe these effects in the most optimistic scenario of hybrid inflation,
revealing the interacting nature of the inflaton field. 
\end{abstract}

\keywords{Primordial Power Spectrum, Trispectrum, Inflationary perturbations, Radiative corrections, 
Renormalization in curved spacetime, Hybrid Inflation} 
\section{Introduction}

Inflation~\cite{Guth:1980zm, Albrecht:1982wi, Linde:1981mu} is one of the most successful paradigms in cosmology: 
it predicts a nearly scale-invariant spectrum of primordial fluctuations~\cite{Mukhanov:1981xt,Mukhanov:1990me}, 
that can act as seeds for the subsequent structure formation, and that is fully confirmed by the measurements of 
the CMB temperature fluctuations. Present results are consistent with simple single-field 
inflationary predictions~\cite{Ade:2015lrj}, and the observed values of the parameters can be realized in 
many different models within particle physics~\cite{Lyth:1998xn, Martin:2013tda}.

While it is quite amazing that we are able to trace back all the Universe Large Scale structure to tiny quantum
fluctuations, the quantum nature itself of the inflaton field puts a limit on the scale-invariance for an interacting
field. Indeed we expect that in many cases higher order corrections may bring new effects into play like 
non-gaussianities~\cite{Maldacena:2002vr,Bartolo:2004if} or features in the primordial spectrum~\cite{Chluba:2015bqa}.
While many of these effects have been discussed and computed in the past for different 
models, see e.g. ~\cite{Starobinsky:1992ts, Martin:2004yi, Hunt:2004vt, Kawasaki:2004pi, Alabidi:2006hg, Covi:2006ci, 
Hamann:2007pa, Chen:2011zf}, 
less attention has been given to the issue of the time-dependence of the intrinsic corrections due to the quantum nature of
an interacting field. Indeed as both the choice of an initial state and the background dynamics break time-translation 
invariance, we may expect that, even in the ``vanilla'' interacting models, the corrections to the
inflaton field correlators contain finite time-dependent contributions. The main goal of this paper is to revisit
the issue of time-dependent corrections for the simplest models of inflation assuming the standard
Bunch-Davies vacuum as initial condition. We will compute and study these terms and disentangle the source 
of the time-dependence and its effect on the inflationary observables. Of course in order to discuss {\it finite} 
corrections, we will also need to address the issue of the renormalization of the correlators to absorb the UV 
divergences into counter-terms~\cite{Ringwald:1986wf,Boyanovsky:1993xf, Boyanovsky:1997cr, Boyanovsky:1996rw, 
Pinter:1999au,Prange:1997iy,Baacke:1999nq,Weinberg:2005vy, Weinberg:2006ac,Bilandzic:2007nb,Borsanyi:2008ar, Senatore:2009cf, 
Baacke:2010bm, Markkanen:2013nwa,Herranen:2013raa,Dresti:2013kya, Hack:2015zwa, Gere:2015qsa}, 
as well as find a regulator for the IR divergences in a quasi de Sitter background~\cite{Seery:2010kh}.

We concentrate here our study on the simple $ \lambda \phi^4 $ model~\cite{Linde:1983gd}, which may 
not be a realistic inflationary model, as it is under siege by the present bounds on the tensor-to-scalar 
ratio~\cite{Ade:2015lrj}, but on the other hand is one of the most studied quantum field theory models and allows for 
comparison with previous literature. Moreover we consider as well the case of hybrid inflation~\cite{Linde:1993cn}
where the interaction with a second massive field is needed for ending the inflationary phase and cannot be
negligible. We expect though the effects found here to be present for any field-theoretic model and to be a 
generic qualitative characteristic of inflation for any interacting field.

We apply the Closed-Time-Path (CTP) formalism to study first the corrections to the power spectrum, 
which arise at the one-loop level, both for nearly massless virtual fields, as the inflaton itself, and for massive 
virtual fields, as it may happen in hybrid inflationary models. In the latter we compute the loop integral with the
full massive propagator in de Sitter for the first time, exploiting the WKB approximation to compute the necessary
counter-terms analytically. Finally we discuss as well the corrections to the 
trispectrum~\cite{Seery:2006vu, Byrnes:2006vq}, where we will see that the time-dependence can arise already 
at the tree-level.

While  our interest is focused on the comprehension of cosmological perturbations of the inflaton field 
as a quantum field on a curved background (see~\cite{Birrell:1982ix,Parker:2009uva} for a discussion of
QFT on curved backgrounds), we investigate explicitly the size and shape of the corrections in this type 
of models and compare them with present bounds.
Indeed, the study of features in the power spectrum and bi/trispectrum is a powerful tool not
only to discriminate between different inflationary models, but also possibly to confirm the quantum 
nature of the inflation field.

After a brief introduction to the \emph{in-in} formalism and to the chosen inflationary models, we will present our 
computation of the radiative corrections to the power spectrum first in Minkowski and then in a quasi-de Sitter spacetime. 
 As expected, we found that the broken Poincar\'e symmetry induced by the expansion of the universe introduces 
 time-dependent corrections to the tree-level, coming from the background, from the renormalization freedom and 
 from the initial state. Finally we will briefly discuss the contributions to the trispectrum and conclude.

\section{Closed-Time-Path formalism}\label{sec:CTP}

In this section we explore the formalism used in cosmology to predict expectation values of quantum observables for 
time-dependent setups. In a Lorentz-invariant quantum field theory one can break explicitly the Poincar\`e symmetry by
 considering for example systems governed by a time-dependent Hamiltonian or a time-dependent background, as in 
an expanding universe.

In our case we are interested in cosmological observables on a Friedmann Lem\ae tre Robertson Walker expanding universe. The metric has the simple form
\begin{equation}
d s^2 = d t^2 - a^2(t)\left[\frac{d r^2}{1- \kappa r^2}+r^2 d \theta^2 + r^2 \sin^2\theta d\phi^2\right],
\end{equation}
where $(r, \theta, \phi)$ are the comoving spherical coordinates, $t$ is the physical time, $a(t)$ is the scale factor and 
$\kappa$ is a constant that can be chosen to be 1, 0 or -1 for a space with positive, zero or negative spatial curvature. 
We will consider flat space and take $ \kappa=0 $ in the following, so that we can rely on the usual Fourier mode expansion for the dependence on the spatial coordinates.
In our analysis the metric will be considered as a classical background. It is clear that the time-dependent scale
factor $ a(t)$ breaks the Poincar\`e symmetry of our system and we therefore cannot rely on obtaining
time-independent radiative corrections as is the case in Minkowski space. 
Often it is useful to write the metric in conformal time as (for $ \kappa =0$)
\begin{equation}
d s^2 = a^2(\tau) \left[ d \tau^2 - d r^2- r^2 d \theta^2 - r^2 \sin^2\theta d\phi^2\right],
\end{equation}
where we define
\begin{equation}
d\tau = \frac{dt}{a(t)}  \quad\Rightarrow\quad  \tau = \tau_\text{in} + \int_{t_\text{in}}^t  \frac{dt}{a(t)}
\end{equation}
which in de Sitter gives simply $ \tau = - \frac{1}{a(\tau) H} $ or $ a(\tau) = - \frac{1}{H \tau} $,
where $ H$ is the constant Hubble parameter and $ \tau_\text{in} $ has been chosen such that $ t = \infty $ 
corresponds to $ \tau = 0 $. So the conformal time is always negative.

We base our computations on \emph{in-in} or CTP formalism \cite{Keldysh:1964ud,Schwinger:1960qe,Calzetta:1986cq,Collins:2012nq}, 
where we consider our system as being described by a time-dependent Hamiltonian $H(t)$ in a state defined by the density matrix $\rho(t)$. 
The expectation value of an operator $\mathcal{O}$ at time $t>t_\text{in}$ is then given by
 
 \begin{equation}
\left<\mathcal{O}(t)\right> = {\rm Tr} \left(\rho(t)\mathcal{O}(t)\right).
\end{equation}

In the interaction picture the time-evolution of the density matrix is described by the Liouville equation
\begin{equation}\label{eq:Liouville}
\begin{cases}
i\frac{\partial \rho(t)}{\partial t} = \left[\hat{H}_I(t), \rho(t)\right] \\
\rho(t_\text{in}) = \rho_\text{in}
\end{cases},
\end{equation}
where we split the Hamiltonian in the free and interacting part $\hat{H}(t)=\hat{H}_0(t)+\hat{H}_I(t)$. 
The solution is given in terms of the time-evolution operator
\begin{equation*}
U_I(t,t_\text{in})={\rm T} e^{-i\int_{t_\text{in}}^t d\tau \hat{H}_I(\tau)},
\end{equation*}
and reads
\begin{equation}\label{eq:rho(t)}
\rho(t)=U_I(t,t_\text{in})\rho_\text{in}U_I^\dagger(t,t_\text{in}) .
\end{equation} 

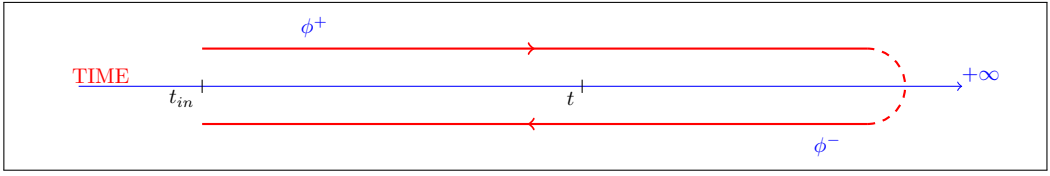
\begin{figure}[tbp]
\centering
\fbox{\parbox[c][2cm]{0.9\textwidth}{\qquad
\begin{tikzpicture}[scale=2.5]
   \input{CTP_profile.tikz}
\end{tikzpicture}}}
\caption{Closed-time contour $\mathcal{C}$ that represents the time-ordering $T_\mathcal{C}$.}
\label{fig:contour}
\end{figure}

Now we have an explicit expression for $\left<\mathcal{O}(t)\right>$
\begin{equation*}
\left<\mathcal{O}(t)\right> ={\rm Tr}\left\{\rho_\text{in} \left({\rm T} e^{-i\int_{t_\text{in}}^t d\tau \hat{H}_I(\tau)}\right)^\dagger \mathcal{O}(t)\left({\rm T} e^{-i\int_{t_\text{in}}^t d\tau \hat{H}_I(\tau)}\right)\right\}
\end{equation*}
that can be understood as the time-evolution from an initial time $t_\text{in}$, where the initial state is given, up to time $t$, where the observable $\mathcal{O}$ has to be evaluated. Then the system evolves back to the initial time. To simplify computations one can extend the time-evolution from time $t$ to infinity
\begin{equation}
\left<\mathcal{O}(t)\right> ={\rm Tr}\left\{\rho_\text{in} \left({\rm T} e^{-i\int_{t_\text{in}}^\infty d\tau \hat{H}_I(\tau)}\right)^\dagger \left({\rm T} e^{-i\int_{t}^\infty d\tau \hat{H}_I(\tau)}\right) \mathcal{O}(t)\left({\rm T} e^{-i\int_{t_\text{in}}^t d\tau \hat{H}_I(\tau)}\right)\right\}.
\end{equation}
From these expression we can apply perturbation theory and define Feynman rules for the
computation of perturbative corrections. Indeed one can define the time-ordering $T_\mathcal{C}$
along the time-path given in Figure~\ref{fig:contour}
and split the field $\phi$ in two components $\phi^\pm$, where the $+$ component propagates along the upper part of the contour and is governed by the Hamiltonian $\hat{H}_I^+(t)= \hat{H}_I [\phi^+]$ and \mbox{the 
$-$} component propagates in the lower part (and is governed by $\hat{H}_I^-(t)= \hat{H}_I [\phi^-]$).

Then one can write the expectation value as
\begin{equation}
\left<\mathcal{O}(t)\right> ={\rm Tr}\left\{\rho_\text{in} {\rm T}_\mathcal{C}\left[\mathcal{O}(t) e^{-i\int_{t_\text{in}}^{+\infty} d\tau [\hat{H}^+_I(\tau)-\hat{H}^-_I(\tau)}\right]\right\}.
\end{equation}
Now, since the two components propagate on the two different parts of the contour, between $ t_{\rm in} $ and
$ \infty $ or opposite, we can use the traditional Feynman rules of the \emph{in-out} approach to treat the expectation 
values perturbatively. The Feynman rules for the models we consider are given in the Appendix~\ref{ap:FeynmanRules}.
In this framework there are four possible contractions of the $\phi$-components and therefore four propagators:
\begin{equation}
G^{\pm\pm}(x, y) = i\left<{\rm T}_\mathcal{C}\left(\phi^\pm(x) \phi^\pm(y)\right)\right>,
\end{equation}
or more explicitly
\begin{align}
G^{+-}(x,y)&=i \langle\phi(y)\phi(x)\rangle,\\
G^{-+}(x,y) &=i \langle\phi(x)\phi(y)\rangle,\\
G^{++}(x,y) &=  \theta(x^0-y^0)G^{-+}(x,y)+ \theta(y^0-x^0)G^{+-}(x,y), \\
G^{--}(x,y) &=\theta(x^0-y^0)G^{+-}(x,y)+ \theta(y^0-x^0)G^{-+}(x,y).
\end{align}
The 4 propagators are not independent. This reflects the fact that the two components should satisfy the boundary condition $\phi^+(\infty)=\phi^-(\infty)$. They are connected through the simple relation
\begin{equation*}
G^{++}(x,y) + G^{--}(x,y) = G^{+-}(x,y) + G^{-+}(x,y).
\end{equation*}
We can regroup them together in a matrix form
\begin{equation}
\mathbf{G}(x,y) = 
\begin{pmatrix}
G^{++}(x,y) & G^{+-}(x,y)\\
G^{-+}(x,y) & G^{--}(x,y)
\end{pmatrix} .
\end{equation}

Since the components of the field $\phi$ are not two independent degrees of freedom, we can transform $\phi^+$ and $\phi^-$ into a more convenient basis. We define $\mathbf{R}$ as
\begin{equation}\label{eq:change_of_basis}
\mathbf{R} = 
\begin{pmatrix}
1/2&1/2 \\
1&-1
\end{pmatrix}.
\end{equation}
The new fields $\phi^{(1)},\phi^{(2)}$ and the new propagators $\mathbf{G_R}$ are given by
\begin{equation}
\begin{pmatrix}
 \phi^{(1)}\\
\phi^{(2)}
\end{pmatrix} = 
\mathbf{R}
\begin{pmatrix}
 \phi^+ \\
 \phi^-
\end{pmatrix}
=
\begin{pmatrix}
( \phi^+ + \phi^-)/2 \\
\phi^+ -  \phi^-
\end{pmatrix}
\end{equation}
and
\begin{equation}
\mathbf{G_R} = \mathbf{R}\,\mathbf{G}\,\mathbf{R}^T =: 
\begin{pmatrix}
i F&G^R \\
G^A&0
\end{pmatrix}.
\end{equation}

The new basis is called the Schwinger basis and in this basis the $\phi^{(2)}$-$\phi^{(2)}$ contraction is vanishing. 
We recognize $G^R$ and $G^A$ as the retarded and advanced propagator and $F$ as the Schwinger 
function~\footnote{In \cite{Fulling:1989nb} the Schwinger function $-iG^{(1)}$ (also named Hadamard function) is 
defined as $-2i F$.}.
\begin{align}
F(x,y) &= -\frac{i}{2}\left(G^{-+}(x, y) + G^{+-}(x, y)  \right),\\
G^R(x, y) &= \theta\left(x_0-y_0\right) \left(G^{-+}(x, y) - G^{+-}(x, y)  \right),\\
G^A(x, y) &=G^R(y, x).
\end{align}
Finally the three abovementioned propagators are connected to the familiar causal (Feynman) propagator $G^F$
\begin{align}
G^F(x,y)=i \left< {\rm T} \left[ \phi(x)\phi(y) \right]\right> &=\frac{1}{2} (G^R(x,y)+G^A(x,y)) +i F (x,y).
\end{align}

Since we consider a spatially flat FRW metric, it is convenient to Fourier transform the propagators in the spatial coordinates and get a function of the momentum $k$ and two times $G^{A/R}(k, t_1,t_2)$ or $ F(k, t_1,t_2) $.
Here we are exploiting the invariance under space translations that insures $G^{A/R} (\vec{x}_1, t_1, \vec{x}_2,t_2)
= G^{A/R} (\vec{x}_1- \vec{x}_2, t_1,t_2)$ . 

In our analysis we are particularly interested in de Sitter spacetime. In this case, for small masses, the propagators 
in momentum space are given in terms of Bessel functions $J_\nu$\cite{Boyanovsky:1996ab,Boyanovsky:2005sh,Boyanovsky:2004ph}
\begin{align}
\label{eq_dSp1}G^{-+}(k, \tau_1,  \tau_2) &=\frac{H^2\pi J_\nu(z)J_{-\nu}(z')}{2\sin(\pi\nu)}( \tau_1 \tau_2)^{3/2}, \\
\label{eq_dSp2}G^{+-}(k,  \tau_1,  \tau_2) &=\frac{H^2\pi J_{-\nu}(z)J_{\nu}(z')}{2\sin(\pi\nu)}( \tau_1 \tau_2)^{3/2},
\end{align}
where $\tau_{1,2}$ is the conformal time, $z=-k \tau_1$, $z'=-k \tau_2$,  and
\[
\nu = \sqrt{\frac{9}{4}-\frac{m^2}{H^2}}, \qquad \text{where} \; m \text{ is the field's mass}.
\]

Assuming a massless field greatly simplifies the propagators, that are now given in terms of the 
Hankel functions $H_\frac{3}{2}^{(1)}, H_\frac{3}{2}^{(2)}$ as
\begin{align}
\label{eq_dS_Hankel1} G^{-+}(k,  \tau_1,  \tau_2) &=\frac{i\pi H^2}{4}  H_{3/2}^{(1)}(z)H_{3/2}^{(2)}(z') \; ( \tau_1 \tau_2)^{3/2}, \\
\label{eq_dS_Hankel2}G^{+-}(k,  \tau_1,  \tau_2) &=\frac{i\pi H^2}{4} H_{3/2}^{(2)}(z)H_{3/2}^{(1)}(z')\; ( \tau_1 \tau_2)^{3/2},
\end{align}
where
\begin{align*}
H_\frac{3}{2}^{(1)}(z) &=\sqrt{\frac{2}{\pi z}}e^{iz}\left(\frac{1}{iz}-1 \right), \\
H_\frac{3}{2}^{(2)}(z) &=\bar{H}_\frac{3}{2}^{(1)}(z).
\end{align*}
Another convenient form for the propagator is given in terms of the Hypergeometric function in position space, 
which is the exact form of the propagators for a massive field in de Sitter spacetime\cite{Bunch:1978yq}: 
\begin{align}\label{eq:hyper}
G^{-+}(\tau_1,\tau_2,x_1,&x_2) = \frac{H^2 \Gamma(3/2 - \nu)\Gamma(3/2 + \nu)}{4 \pi^2}  \,_2F_1\left(3/2 - \nu,3/2 + \nu,2,1-r/4 \right), \\ 
\nu&=\sqrt{\frac{9}{4}- \frac{m^2}{H^2}}, \quad r= \frac{\left(-(\tau_1-\tau_2)^2+|\mathbf{x}_1-\mathbf{x}_2|^2 \right)}{\tau_1\tau_2}.
\end{align}

\begin{figure}[tbp]\label{hyp}
	\centering\includegraphics[width=.6\textwidth]{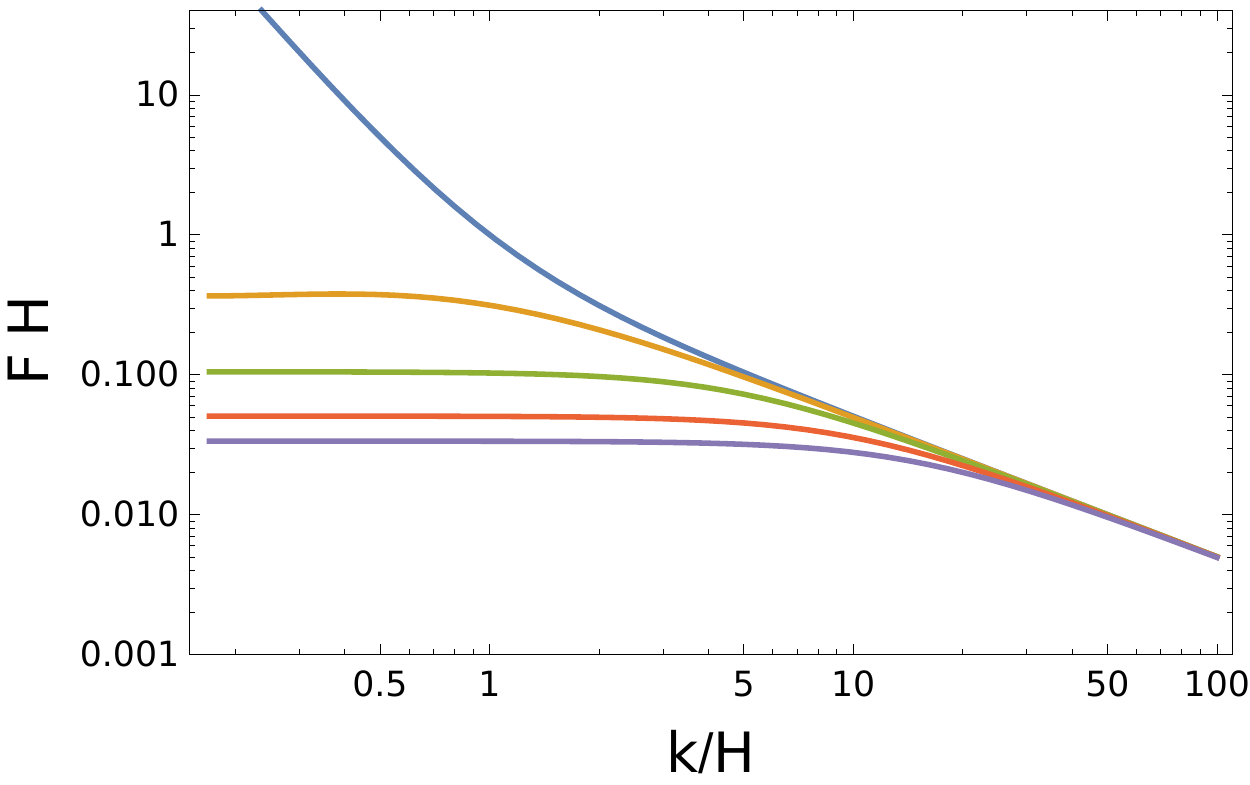}
	\caption{Hadamard propagator derived from the Hypergeometric (full massive) function for different masses: the yellow curve 
	represents $m = 2H$, the green curve $m = 5H$, the red curve $m = 10H$ and the purple curve $m=15H$. The blue curve is the massless propagator given by the Hankel function with $\nu = 3/2$.}
\end{figure}

As there is no way to obtain an analytical form of the Fourier transformation of this propagator,
we take the numerical value for the hypergeometric function and compute the massive propagator 
in Fourier space with numerical methods with Mathematica \cite{Mathematica}. 
The integrations needed for the Fourier transformation are not always numerically stable, but 
they are well-behaved for masses $ m > H $, which is the region of interest in our computations.
In Figure~\ref{hyp} we show the behaviour of the Hadamard propagator reconstructed from the Fourier transformed full massive Hypergeometric function
for different masses and compare it with the massless propagator, which shows explicitly 
an IR-divergence for $ k=0 $.

While the massive propagator has a much better behaviour in the infrared regime and avoids completely IR divergences 
in the computations, the price to pay for this procedure is that the expression in eq.~(\ref{eq:hyper}) is an implicit
function of the coordinates.  In particular the computation of explicit renormalisation counter-terms in
the Fourier space with the full massive propagator is difficult. For this purpose, we will introduce the
approximated propagator in the semiclassical WKB limit.

For a very massive quantum field with a Bunch-Davies vacuum \cite{Allen:1985ux} as initial state, a good solution of the
free field equations can be obtained in the WKB approximation. 
Let's define the Hadamard propagator from the field solution $V(k,\tau)$
\[
F(k, \tau_1, \tau_2) =\mathcal{R}\left[V(k,\tau_1)V^*(k,\tau_2)\right].
\]
Then in the WKB limit $H/m \ll 1$ one can approximate $V$ as\cite{Jackson:2010cw, Jackson:2011qg,Jackson:2012fu}
\begin{equation}
V(k,\tau)=\frac{{\rm exp}\left[-i \int_{\tau_\text{in}}^\tau d\tau_1 \, \sqrt{k^2 + m^2 a(\tau_1)^2}  \right]}{\sqrt{2}a(\tau)\left(k^2 + m^2 a(\tau)^2  \right)^{1/4}}.
\end{equation}
From this we can define the WKB Hadamard propagator
\begin{equation}\label{eq:WKBprop}
F_{WKB} (k, \tau_1, \tau_2) =\frac{\cos\left[ \int_{\tau_1}^{\tau_2} d\tau \, \sqrt{k^2 + m^2 a(\tau)^2}\right]}{2a(\tau_1)a(\tau_2)\left(k^2 + m^2 a(\tau_1)^2  \right)^{1/4}\left(k^2 + m^2 a(\tau_2)^2  \right)^{1/4}}.
\end{equation}
Note that this propagator is suppressed by $ m^{-1} a^{-3} (\tau_1) $ for $ \tau_1 =\tau_2 $ and $ k \ll m a(\tau_1) $ \cite{Weinberg:2006ac}, 
while in the UV the dependence is weaker $ \propto a^{-2} (\tau_1) $, exactly as the subleading piece of the massless propagator.
This propagator provides with a very good approximation of the full massive propagator
in eq.~(\ref{eq:hyper}) in the UV  and for $ m \gg H $ and we will therefore exploit it for the 
analytic evaluation of the counter-terms.

\begin{figure}[tbp]
	\centering\includegraphics[width=.6\textwidth]{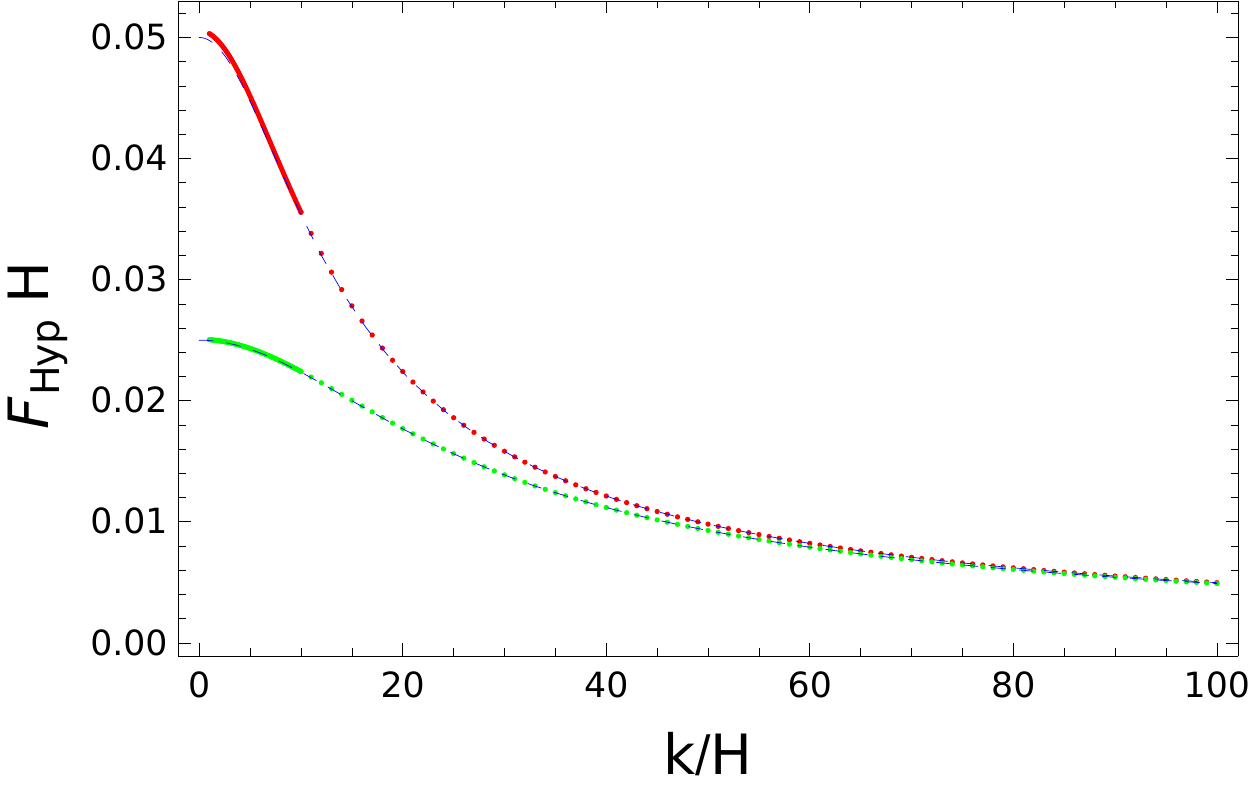}
	\caption{Comparison of the massive Hadamard propagator reconstructed from the Hypergeometric function (\ref{eq:hyper}) (list of points) and the WKB propagator (\ref{eq:WKBprop}) (dashed blue lines) for a mass of $10H$ (red points) and $20H$ (green points).}
	\label{fig:hypWKB}
\end{figure}

In Figure~\ref{fig:hypWKB} we plot both the Fourier transform of the full massive propagator
on discrete points for $ m > H $ and compare it with the WKB expression. We see that
the two propagators overlap perfectly.

\begin{figure}[tbp]
	\centering\includegraphics[width=.6\textwidth]{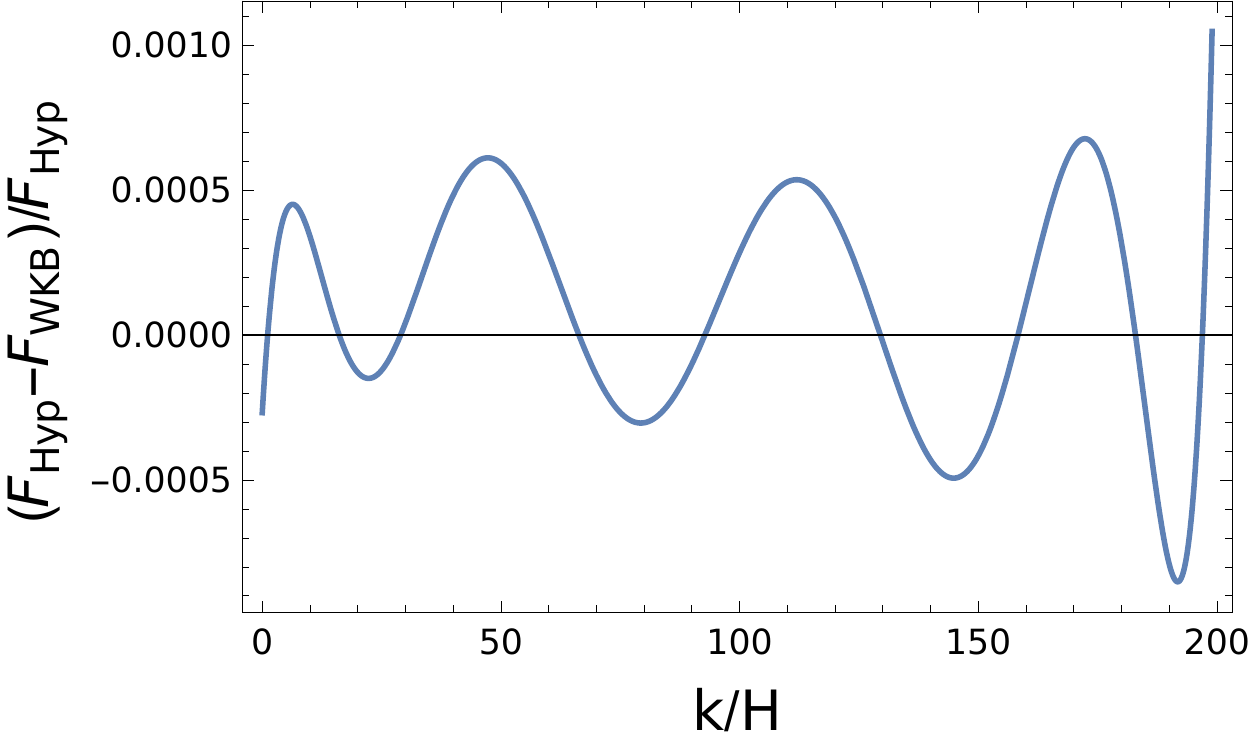}
	\caption{Relative difference between the Hypergeometric and the WKB propagator for 
                 a scalar field of mass $60H$.}
	\label{fig:hypWKBcomparison}
\end{figure}

Since it is important to be able to subtract the complete UV divergence in the loop diagrams,
we also explore the precision of the approximation and give in Figure~\ref{fig:hypWKBcomparison} 
the relative difference between the Hypergeometric and the WKB propagator for a mass of $60H$. 
The reason for the oscillation in the difference and the apparent increase at large $k$ is due to the 
fact that we are fitting the Fourier transformed Hypergeometric function at discrete points with a 
polynomial function exactly on this range. At the boundary the fit becomes less reliable (and we will 
exclude the last part of that range on our analysis). Nevertheless, we verified for different masses that 
the integrated difference, which is crucial for the cancellation of the UV divergence in the loops, is always 
finite and less than 0.001\%.

In our analysis we compute corrections to the primordial spectrum coming from massive and massless fields.  
Since we are interested in the finite corrections to the field correlators, we have first to renormalize our
quantities and subtract the UV and IR divergences in the minimal subtraction (MS) renormalization scheme.  
For each case we look for the best strategy to compute the MS counter-terms analytically, employing 
the simplest propagator with the same UV divergence.

 In the case of very light virtual fields, like the inflaton field, we use the Hankel propagators in the loop
with an index $ \nu = 3/2 - \varepsilon $, where $ \varepsilon  $ takes into account a non-vanishing mass and 
contributes to the tilt in the CMB spectra. We will see that $ \varepsilon $ will act as IR-regulator as well.
Instead in the case of a heavy field in the loop with $ m \gg H $, we use the exact propagator given in terms 
of the hypergeometric function in the loop computation. Since it is hard to extract the UV dependence numerically, 
we subtract the UV divergence with the counter-term obtained from the WKB approximation.
Note of course that in this latter case the loop computation is IR-finite.

\section{Inflationary models and the power spectrum}

Let us consider a simple model of inflation consisting in a scalar field $\phi$ with canonical kinetic term and
with a potential $V(\phi)$.  
The Lagrangian density for such a scalar theory is given by 
\begin{equation}\label{eq:lagrang}
\mathcal{L}[\phi] = \sqrt{-g}\left( \frac{1}{2} \partial_\mu \phi \partial^\mu \phi -\frac{1}{2}m^2\phi^2- V(\phi) 
+ \frac{\xi}{2}\; \mathcal{R}\; \phi^2 \right)+ \delta\mathcal{L} ,
\end{equation}
where the metric $g_{\mu\nu}$ has signature $(+---)$, $g=\det(g_{\mu\nu})$ {and the counter-terms $\delta\mathcal{L}$ are}
\begin{equation}
\delta\mathcal{L} =\sqrt{-g}\left( \frac{1}{2}\delta Z\; \partial_\mu \phi \partial^\mu \phi -\frac{1}{2}\delta m^2\; \phi^2- \delta V (\phi) 
+ \frac{\delta \xi}{2} \;\mathcal{R} \; \phi^2 
 \right).
\end{equation} 
We will consider in the following that the curvature coupling $ \xi  $ is vanishing or very small at tree-level,
so that it does not affect the inflationary dynamics, contrary to what happens in Higgs inflation~\cite{Bezrukov:2007ep}.
Note that even for vanishing $ \xi $, a contribution $ \delta \xi $ will be automatically generated at the one-loop level, 
as we will see later.
Then slow-roll inflation takes place if the potential $V(\phi)$ satisfies the slow-roll  conditions~\cite{Lyth:1998xn}
\begin{align*}
\epsilon = \frac{M_P^2}{2} \left(\frac{V'}{V}  \right)^2 \ll 1, \quad\quad
|\eta| = M_P^2 \left|\frac{V''}{V}  \right| \ll 1,
\end{align*}
where $ M_P = 2.4 \,\times 10^{18}\,$GeV is the reduced Planck mass. 
In the slow-roll phase, the classical dynamics of the field is reduced to the attractor solution, 
given by the simplified equation of motion,  
\begin{equation}
\ddot \phi + 3 H \dot \phi + V'(\phi) \sim 3 H \dot \phi + V'(\phi)  = 0
\quad \rightarrow \quad \dot\phi = - \frac{V'(\phi)}{3 H} \; ,
\end{equation}
so that the number of e-folding from the classical field value $ \phi $ to the end of
inflation at $ \phi_\text{end} $  is simply given by
\begin{equation}
N(\phi) = \int_t^{t_\text{end}} dt\; H(t) = \frac{1}{M_P^2} \int_{\phi_\text{end}}^{\phi} d\varphi\; \frac{V(\varphi)}{V'(\varphi)}
\; .
\label{eq:Nphi}
\end{equation}

In the standard inflationary paradigm the quantum fluctuations $\delta \phi$ generates primordial fluctuations of 
the curvature tensor that act as seeds for the temperature anisotropies in the Cosmic Microwave Background 
and of the observed large scale structure of our universe.
In the slow roll approximation one can obtain the power spectrum of the curvature perturbations
in terms of the inflaton potential $V(\phi)$ as
\begin{equation}
\mathcal{P}_{\mathcal{R}}(k) = \frac{1}{12 \pi^2 M_P^6 } \left.\frac{V^3}{V'^2}\right\vert_{k=aH},
\label{eq:P_RvsP_phi}
\end{equation}
where the potential and its first derivative are evaluated at the horizon exit~\footnote{The curvature perturbation $\mathcal{R}$ freezes ($\dot{\mathcal{R}}_k \approx 0$) once the mode crosses the horizon.} $k/a = H$. 
The relation between the comoving scale $ k $ at horizon exit and the corresponding value of the 
classical inflaton field is given simply by eq.~(\ref{eq:Nphi}) as
\begin{equation}
\log\left( \frac{k_\text{end}}{k}  \right) = \log\left( \frac{a (t_\text{end}) H(t_\text{end})}{a(t_k) H(t_k)} \right) \sim N(\phi(k)) \; .
\end{equation}
where here $ k_\text{end} $ is the comoving scale characterized by horizon exit at the end of slow-roll
inflation and we have taken $ H(t_\text{end}) \sim H(t_k) $.
From these relation one can also obtain a prediction for the spectral index of the curvature
perturbations in first-order in the slow-roll parameters as 
\begin{equation}
n_s(k) - 1 = \frac{d \mathcal{P}_{\mathcal{R}}}{d\ln k} =  
2\eta (\phi(k)) -6\epsilon(\phi(k)).
\end{equation}

\subsection{Explicit Models}

After giving the general predictions, let us discuss in detail the specific models that we will consider
in the rest of the paper. The simplest models to realise slow-roll inflation are the class of large field 
models based on monomial potentials. In the following we will study the simple quartic potential,  
\begin{equation}\label{eq:V4}
V(\phi) = \frac{\lambda}{4 !} \phi^4\; ,
\end{equation}
as the most studied model with an interacting inflaton field.
In the chaotic inflation scenario it is assumed that the universe emerged from a quantum gravitational energy 
density comparable to the Planck density. Therefore there is a large friction term in the Friedmann equations 
and the field is slowly rolling down the potential. Inflation ends when the field is of the order of 
the Planck scale, as we have for the slow-roll parameters
\begin{align*}
\epsilon = \frac{8 M_P^2}{\phi^2} \ll 1, \quad\quad
\eta = \frac{12 M_P^2}{\phi^2}  = \frac{3}{2} \epsilon \ll 1,
\end{align*}
giving directly 
\begin{equation}
n_s(k) - 1 = 2\eta-6\epsilon = - 3 \epsilon =  - \frac{24 M_P^2}{\phi^2} = - \frac{3}{N(k)}.
\end{equation}
At the same time the normalization of the power spectrum fixes the scale of the coupling $ \lambda $ as
\begin{align}
\mathcal{P}_{\mathcal{R}}(k_*) &= \frac{1}{12 \pi^2 M_P^6 } \left.\frac{V^3}{V'^2}\right\vert_{k_*=a H}
=\frac{ \lambda\; \phi_*^6}{4608 \pi^2 M_P^6} = \frac{\lambda N_*^3}{9 \pi^2 },
\end{align}
where $ N_* = 3/(1- n_s(k_*)) $ giving
\begin{equation}
\lambda = \frac{\pi^2}{3} (1-n_s(k_*))^3 \; \mathcal{P}_{\mathcal{R}}(k_*) = 2.97 \times 10^{-13} \;  ,
\label{eq:lambda}
\end{equation}
for the observed values of $ \mathcal{P}_{\mathcal{R}}(k_*) = A_s = (2.20 \pm 0.08 ) \times 10^{-9} $ 
and $ n_s(k_*) = 0.9655 \pm 0.0062 $ \cite{Ade:2015lrj}.
We see therefore that the self-coupling of the field has to be very suppressed and the potential
flat enough in order to generate the observed small fluctuations.

But this it not always the case. Indeed in models with more than one scalar field also much larger couplings 
can be present.
This leads us to the second class of models that we will discuss: hybrid inflation. 
The simplest version is given by two scalar fields with the following effective potential
\begin{equation}\label{eq:hybrid}
V(\phi, \sigma)=\frac{m^2}{2}\phi^2+\frac{1}{4 g}(M^2- g \sigma^2)^2 + \lambda_h^2 \phi^2\sigma^2
+ \Delta V_{CW} (\phi),
\end{equation}
where $ \Delta V_{CW} (\phi) $ gives the one-loop correction to the effective potential {\it a la} Coleman-Weinberg~\cite{Coleman:1973jx},
\begin{equation}\label{eq:CWpotential}
\Delta V_{CW} (\phi) = \sum_i (-1)^F 
\frac{m_{i,\text{eff}}^4(\phi)}{64 \pi^2} \left( \ln\left( \frac{m_{i,\text{eff}}^2(\phi)}{\mu^2}  \right) - \frac{3}{2}  \right),
\end{equation}
where $ (-1)^F $ gives a minus sign for fermionic fields, the sum runs over all the field
content of the theory and $ \mu $ is an arbitrary renormalization scale. 
In the case of the potential above, we have then that for $ \sigma = 0 $ the effective mass of $\sigma$ 
is given by $ 2 \lambda_h^2\phi^2-M^2$ and the effective mass of $\phi $ is just $ m^2 $ and so
\begin{equation}\label{eq:CWpotential2}
\Delta V_{CW} (\phi) = 
\frac{( 2 \lambda_h^2 \phi^2 - M^2)^2}{64 \pi^2} \left( \ln\left( \frac{2 \lambda_h^2\phi^2 - M^2}{\mu^2}  \right) - \frac{3}{2} \right)
+ \frac{m^4}{64 \pi^2} \left( \ln\left( \frac{m^2}{\mu^2}  \right) - \frac{3}{2} \right)\; .
\end{equation}
Here the last term due to the loop-corrections from the inflaton itself is constant
as long as $ \sigma = 0$ and does not contribute to the dynamics.

For $\phi$ larger than $\phi_c = M/(\sqrt{2} \lambda_h) $ the only minimum of the potential is at $\sigma=0$. 
Therefore at the beginning of inflation the field $\sigma$ rolls down the potential to $\sigma=0$ while 
$\phi$ remains large and drives inflation. 
When the inflaton field becomes smaller than $M/(\sqrt{2} \lambda_h)$ a phase transition occurs and inflation ends.
Then the two fields rapidly fall to the absolute minimum of the potential at $\phi=0$ and 
$\sigma^2=M^2/g$. 
If the slope of the potential is dominated by the inflaton mass term, one has
\begin{align*}
\epsilon = \frac{2 M_P^2}{\phi^2} \ll 1, \quad\quad
\eta = \frac{2 M_P^2}{\phi^2}  =  \epsilon \ll1 .
\end{align*}

The relation between the scale $ k $ at horizon exit and the corresponding value of the classical inflaton
field is given in this case simply by
\begin{equation}\label{eq:Efolds}
n_s(k) - 1 = 2\eta-6\epsilon = - 4 \epsilon =  - \frac{8 M_P^2}{\phi^2(k)} = - \frac{2}{N(k)}   .
\end{equation}
and the normalization imposes the constraint
\begin{equation}
\frac{m^2}{M_P^2} = \frac{3 \pi^2}{2}  (1-n_s(k_*))^2 \; \mathcal{P}_{\mathcal{R}}(k_*) = 0.38 \times 10^{-10} \;   ,
\end{equation}
giving $ m = 6.22 \times 10^{-6} M_P $ compared to $ H = \sqrt{2 N(k_*)/3}\; m > m$ .

In the case instead where $ m \sim 0 $, the slope can be provided by the one-loop corrections related to
the coupling $ \lambda_h $. Indeed from the first derivative of the potential, we see that the one-loop corrections 
dominate for 
\begin{equation}
\lambda_h^2 \geq 2\pi \frac{m}{\phi}  \; .
\label{eq:lambdahsmall}
\end{equation}
Models of inflation where the inflationary potential is dominated by the quantum corrections have been studied
especially in supersymmetry~\cite{Dvali:1994ms, Dvali:1997uq} and supergravity~\cite{Buchmuller:2000zm}. 
In that case inflation can be obtained also for value of the quartic coupling $\lambda_{h} \sim 0.001 $ and we will use this 
value as a maximal value~\footnote{It is well-known that in Higgs inflation, the quartic coupling can be even larger, 
but in that case the effective coupling for the canonically normalized field is $ \lambda/\xi^2 $, which due to the 
power spectrum normalization is of order $ \sim 10^{-9} $, not much larger than for the classic chaotic inflation.}.
Indeed in the following, we will investigate the radiative corrections to the power spectrum for a hybrid model inspired by 
a supersymmetric theory with the approximate scalar potential~\cite{Buchmuller:2000zm}
\begin{eqnarray}
V &=& \lambda_h^2 |M_G^2 - \Sigma^2|^2 + 4\lambda_h^2|\Phi |^2|\Sigma|^2 + M_S^4 + \Delta V_{CW}
\label{eq:suppot} \\
&=& \lambda_h^2 M_G^4 + \frac{\lambda_h^4 M_G^4}{8\pi^2} \left[ \ln \left( \frac{2\lambda_h^2 \phi^2}{\mu^2} \right) +
{\cal O} \left( \frac{M_G^4}{\phi^4} \right) \right],
\end{eqnarray}
where $\Phi, \Sigma$ are complex scalar fields,  $M_S$ is the supersymmetry breaking scale and
$ \phi $ denotes the real part of $\Phi$.  In the last line we have taken $ \Sigma = 0 $ to obtain the potential during 
the inflationary phase and identified the inflaton with the real part of $ \Phi$. 
We see that in this case the slow-roll parameters are then
\begin{align*}
\epsilon = \frac{\lambda_h^4}{32\pi^4} \frac{M_P^2}{\phi^2}, \quad\quad
\eta =  - \frac{\lambda_h^2}{4\pi^2} \frac{M_P^2}{\phi^2},
\end{align*}
both small for small enough $ \lambda_h $ or large enough $ \phi $. So the spectral index is given by
\begin{equation}
n_s(\phi) - 1 = 2 \eta = - \frac{\lambda_h^2}{2\pi^2} \frac{M_P^2}{\phi^2} = 
- \left(N + \frac{2\pi^2}{\lambda_h^2} \frac{M_G^2}{M_P^2}\right)^{-1}.
\end{equation}

In the region of large $ \lambda_h $, which is where
the one-loop corrections are most important, the spectrum normalization just fixes the scale $ M_G $ since we
have
\begin{align}
\mathcal{P}_{\mathcal{R}}(k_*) &= \frac{1}{12 \pi^2 M_P^6 } \left.\frac{V^3}{V'^2}\right\vert_{k_*=a H}
=\frac{4 \pi^2  M_G^4\; \phi_*^2}{3 \lambda_h^2 M_P^6} = \frac{2 N_* M_G^4}{3 M_P^4} + \frac{4 \pi^2 M_G^6}{3 \lambda_h^2 M_P^6},
\end{align}
where the second term is negligible. Taking then $ N_* = 1/(1- n_s(k_*)) $ we obtain
\begin{equation}
\frac{M_G}{M_P} = \left(\frac{3}{2} (1-n_s(k_*)) \; \mathcal{P}_{\mathcal{R}}(k_*) \right)^{1/4} = 3.26 \times 10^{-3} \;  ,
\label{eq:MG-normalization}
\end{equation}
compatible with the Grand Unification scale. We see that in this case both the spectral index and the spectrum normalization
are practically independent from the coupling, so that it can be chosen large. Requiring though that the slow-roll
conditions are satisfied until the critical point $ \phi_c = M_G $, we have from the condition $ |\eta | < 0.1 $, the
maximal value of the coupling as
\begin{equation}
\lambda_h < 2\pi \frac{M_G}{\sqrt{10}\; M_P} = 6.8 \times 10^{-3}\; .
\label{eq:lambda_h_max}
\end{equation}

Finally, we will discuss also  an example of a massive spectator field, with a potential  similar to the hybrid model in eq.~(\ref{eq:hybrid}),
but with no symmetry breaking along the spectator direction.  In this case the spectator's mass $m_\sigma$ is constrained to be 
smaller than $ 5 \times 10^{-2} M_P$ and the coupling $ \lambda_h^2 < 10^{-5} $ so that the spectator's one-loop corrections do 
not modify the inflationary dynamics, which we consider determined by the $ m^2 \phi^2 $ term. So in this case the coupling takes 
an intermediate value between the quartic inflation case and the hybrid inflation case, but the mass $ m_\sigma $ is in general 
even larger than $ M_G$. 

In the following we will use these three models, non-supersymmetric and supersymmetric hybrid inflation and quadratic
inflation with a spectator field, as benchmark models for how large the quantum corrections from a massive field can be. 
Note that the hybrid inflation case is indeed the most optimistic one, since the hybrid field $ \sigma $ must couple sufficiently 
strongly to the inflaton to be kept vanishing during the inflationary phase.

\subsection{One-loop corrections to the Primordial Power Spectrum}

The power spectrum of primordial curvature perturbations is an important tool for distinguishing among different 
inflationary models. Indeed, as we have seen, the spectral index depends on the inflationary potential.
Apart for the departure from near-scale invariance, also features in the primordial spectrum are smoking
gun signals for particular models of inflation. They are predicted at tree-level for extended models~\cite{Chluba:2015bqa, Ade:2015lrj}
or for non-standard initial states \cite{Brandenberger:2012aj}
but they appear here in any case at the one-loop level.
In this paper we want to investigate the possible features in the primordial spectrum due to the time-dependence
coming from radiative corrections.

In the case of single field inflation the power spectrum of the curvature perturbations is directly connected to the 
correlation functions of the inflaton fluctuations.
Let $\delta\phi (t, x) $ be the quantum fluctuations around the classical inflaton field $ \phi(t) $, and let us consider
its Fourier transform with respect to the spatial coordinates.
The power spectrum for the inflaton fluctuations is then defined as
\[
(2\pi)^3\delta^{(3)}(k+k')\mathcal{P}_{\delta\phi}(k)= \left< \delta\phi_k\delta\phi_{k'}\right>,
\]
where $\delta\phi_k$ are the Fourier modes. The power spectrum for the inflaton field in the \emph{in-in} formalism is given by
\begin{equation}\label{eq:PowerSpectrum}
\left.\mathcal{P}_{\delta\phi}(k)= 
 {\rm Tr}\left\{\rho_\text{in} \mathcal{T}_\mathcal{C} \left[ |\delta\phi_k|^2\,e^{-i\int_{t_\text{in}}^{\infty} d \tau \left[\hat{H}_I^+(\tau)-\hat{H}_I^-(\tau)\right]}\right]\right\}\right\vert_{k=a H},
\end{equation}
where the trace is evaluated at the horizon exit in order to have a spectrum that depends only on $k$.

The perturbative expansion has the following diagrammatic representation
\begin{align*}
\mathcal{P}_{\delta\phi}(k) = 
\left(\quad\parbox{18mm}{\vspace{4mm}\input{propag++.tex}}\quad+\quad\parbox{14mm}{\vspace{4mm}
\begin{fmffile}{tadpole}
\begin{fmfgraph}(40,30)\fmfkeep{tadpole}
\fmfset{thick}{0.4mm}
\fmfpen{thick}
\fmfleft{i}
\fmfright{o}
\fmf{plain}{i,v,v,o}
\end{fmfgraph}
\end{fmffile}}\quad+\quad\ldots\quad\right) .
\end{align*}
What is typically extracted from data is the power spectrum of the curvature fluctuations that can 
be computed directly from $\delta\phi$ in the case of single scalar field models as
\begin{equation}\label{eq:PSinf}
\mathcal{P}_{\mathcal{R}}(k)=  \frac{k^3}{4\pi^2} \left(\frac{H^2}{\dot\phi^2}  \right)\mathcal{P}_{\delta\phi}(k)
=  \frac{k^3}{4\pi^2} \frac{1}{2\epsilon\; M_P^2} \mathcal{P}_{\delta\phi}(k).
\end{equation}
where the last expression is given in the slow roll approximation and is equivalent to eq. (\ref{eq:P_RvsP_phi}).

In the following we will discuss corrections to the power spectrum coming from the computation of
the one-loop corrections to the inflaton field correlators and assume that the relation (\ref{eq:PSinf})
still holds also at that order~\footnote{In principle one has to include the one-loop corrections
for the scalar potential in eq.~(\ref{eq:CWpotential}) in eq. (\ref{eq:P_RvsP_phi}), but for the simple 
monomial potential they are numerically negligible \cite{Bilandzic:2007nb} and do not introduce any additional time-dependence, 
so we will neglect them.}.

\section{Renormalization and one-loop corrections in Minkowski spacetime}

Before starting to consider the radiative corrections of the two-point correlation function in quasi-de-Sitter 
spacetime, let us discuss the simpler case of the quantum corrections in Minkowski spacetime. 
This will allow us to obtain the UV counter-terms of the theory, which are universal,
and also to see the effect of the time-dependence of the initial state on a static background.
Indeed it has been proven with algebraic QFT methods that the UV divergences in curved space-time are 
of the same order as in Minkowski and the renormalization freedom (i.e. the number and type of 
counter-terms needed)  is therefore the same~\cite{Brunetti:1997ub,Brunetti:1997ye,Brunetti:1999jn}. 
Moreover the counter-terms should take the form of covariant expressions~\cite{Hollands:2001nf}
and be reabsorbed into redefinitions of the wave function, the mass and coupling constants.

Let us consider a massive scalar field theory with quartic coupling $\frac{\lambda}{4!}\phi^4$. 
The CTP Minkowski propagators are then simply
\begin{align}
F(w_k, t_1, t_2) &= \frac{\cos((t_1-t_2)w_k)}{2 w_k},\\
G^R(w_k, t_1, t_2) &= \theta(t_1-t_2)\frac{\sin((t_1-t_2)w_k)}{w_k},
\end{align}
where $w_k=\sqrt{k^2+m^2}$. The one-loop contribution to the two-point function amplitude is given by
\begin{equation}\label{eq:tadpoleM}
2\int_{t_\text{in}}^t d t_1\, \left(-iG^R(k,t,t_1)\right)F(k,t_1,t)\left[ \left(\frac{-i\lambda}{2} \right) \int \frac{d p^3}{(2\pi)^3} F(p,t_1,t_1)\right], \; 
\end{equation}
where the expression in the square bracket is the loop integral.
By integrating analytically this expression with an explicit UV cut-off $\Lambda $  we obtain
the amputated 2-point function as
\begin{align}
A_{\rm amp}&=\frac{-i\lambda}{8\pi^2}\left(\Lambda^2 \sqrt{1 + \frac{m^2}{\Lambda^2}} -m^2\arcsinh \left(\frac{\Lambda}{m} \right) \right)\\
&= \frac{-i\lambda}{8\pi^2}\left(\Lambda^2 + \frac{1}{2} m^2 + m^2\log\left(\frac{m}{2\Lambda} \right) \right)+\mathcal{O}\left(\Lambda^{-2}\right).
\end{align}

The result is consistent with analogous \emph{in-out} results in the literature (e.g. \cite{Ramond:1981pw}).  
We see here directly both the UV quadratic and logarithmic divergence, regulated by $ \Lambda $, while
the result is IR-finite also in the limit of vanishing mass.
Applying a minimal subtraction scheme to remove the UV divergences, we arrive to the finite expression
\begin{equation}
[A_{\rm amp}]_{\rm ren}= \frac{-i\lambda m^2}{16\pi^2}\left(1+\log\left(\frac{m^2}{4\mu^2} \right) \right),
\end{equation}
where we introduced an arbitrary energy scale $\mu$ and taken the mass counter-term as
\begin{equation}
\delta m^2 =  \frac{\lambda}{16\pi^2}\left(\Lambda^2 - m^2\log\left(\frac{\Lambda}{\mu} \right) \right) \; .
\label{eq:Mink-counterterm}
\end{equation}
By considering the full two-point function with the external propagators we finally have
\begin{equation}
-i [{A_\text{amp}}]_{\rm ren} \cdot \frac{\sin\left(\sqrt{k^2+m^2}(t-t_\text{in})\right)^2}{2\left({k^2+m^2}\right)^{3/2}}.
\end{equation}

\begin{figure}[btp]
	\centering\includegraphics[width=.6\textwidth]{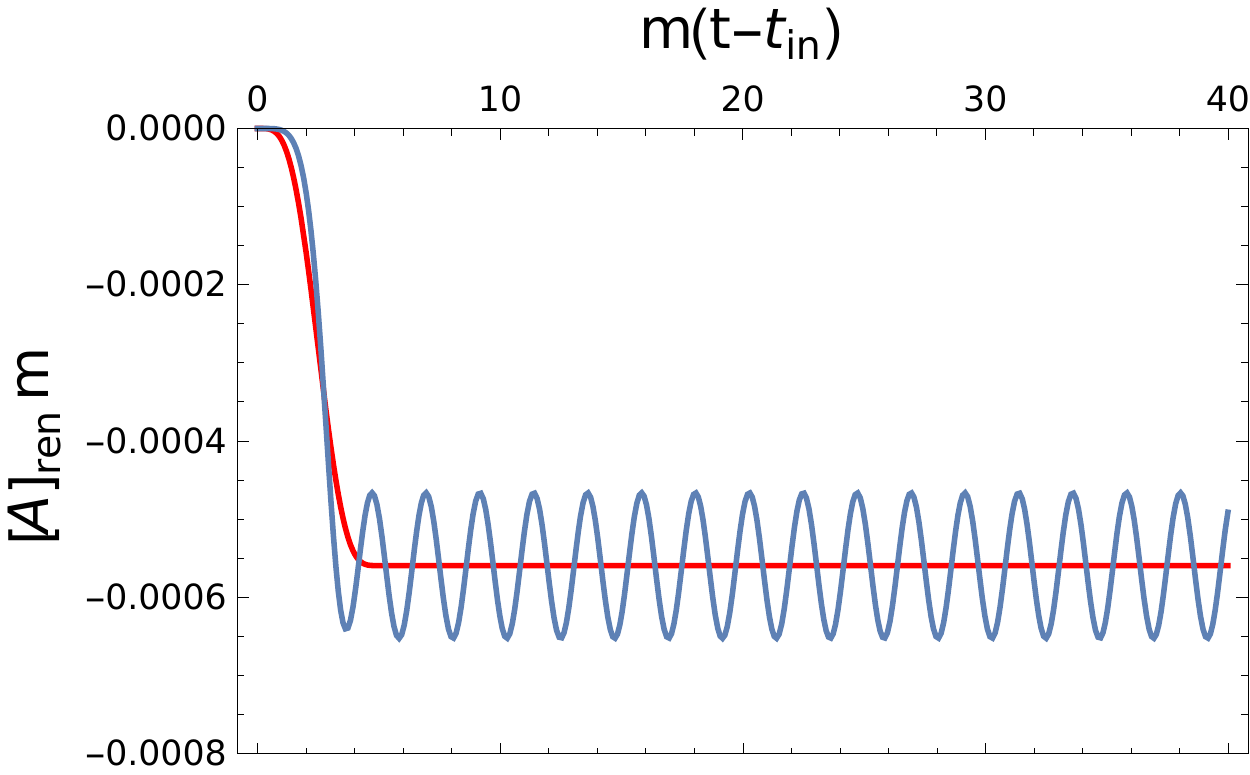}
	\caption{Renormalized two-point function using different continuous interaction profiles $\lambda(t)$ where $\lambda(t-t_{\rm in}<0) = 0$ and $\lambda(t-t_{\rm in}>5/m) = 1$ for $k/m=1$ and $\Lambda/\mu=1$. The blue line corresponds to an interaction profile with a cubic behaviour about time 0 and before the transition. The red line corresponds to the adiabatic switching-on.}
\label{fig:mink_tadpole}
\end{figure}

The final expression for the tadpole is time-dependent, although we are in Minkowski spacetime. This time dependence 
is related to the time-evolution of the interacting theory from the initial time $ t_{\rm in} $, where we have set
abruptly the initial Minkowski vacuum state, and not due to any background evolution. 
In the limit when $ t_{\rm in} \rightarrow \infty $ and the interaction is switched on in an adiabatic way,
we expect this oscillations to disappear (see Figure \ref{fig:mink_tadpole}).

\section{One-loop corrections in quasi de Sitter space from a (nearly) massless field}

We apply the \emph{in-in} formalism summarized in Section \ref{sec:CTP} to the computation of the two-point function
for the inflaton field in quasi-de-Sitter spacetime. 
The scalar field power spectrum at tree-level is simply given by the $F$ propagator 
$\left<{\rm T}_\mathcal{C}\left(\phi^{(1)}(x) \phi^{(1)}(y)\right)\right>$. 
For a pure massless scalar field theory in de Sitter the propagator assumes a simple form
\[
F(k, \tau, \tau) = \frac{H^2}{2k^3} \left(1+k^2\tau^2\right)
\]
and one obtains an exact scale-invariant power spectrum~\footnote{The momentum dependence $k^2 \tau^2$ disappears 
because the power spectrum is evaluated at the horizon exit $k=a H$, i.e. $k\tau  = - 1$ in de Sitter spacetime.}, 
with small deviation due to the $ \tau $ dependence of the background variables $ H(\tau), \dot\phi(\tau)$:
 \begin{equation}\label{eq:treePphi}
P_{\cal R} (k) = \left. \frac{H^4(\tau)}{4\pi^2 \dot\phi^2(\tau)} \right|_{ k\tau  = - 1} .
\end{equation}
For the general case of a massive inflaton,
we consider  a deviation from $ \nu = \frac{3}{2} $ given by a term proportional to $m\ll H$ that will 
regulate our theory in the infrared regime. Indeed expanding the Hankel functions of index 
$ \nu = \sqrt{\frac{9}{4}- \frac{m^2}{H^2}} $ in eqs.~(\ref{eq_dS_Hankel1}), (\ref{eq_dS_Hankel2}) 
for small mass and momentum one gets
\begin{equation}\label{eq:propeps}
F(k, \tau, \tau) = \frac{H^2}{2k^3} \left(k|\tau |\right)^{2\varepsilon},
\end{equation}
 with $\varepsilon = m^2/3H^2$. 
 We will later consider the index $ 2 \varepsilon \sim  n_s -1 $ to estimate the order of
 magnitude of the IR cut-off. Indeed in the case of a quartic potential the inflaton fluctuations
 feel an effective mass term of order $ m_{\rm eff} \sim \lambda \phi^2 $ from the V.E.V. of the classic 
 field and here we will consider such quantity as our IR cut-off. Other possibilities have been
 considered in the past, e.g. in~\cite{Sloth:2006az}, the Hubble scale at the beginning of
 inflation has been taken as the IR cut-off in the one-loop integrals, with the result of
 a very large IR correction, dependent on the duration of inflation ~\cite{Riotto:2008mv}.
 Here we take instead the more conservative view that the slowly changing
 effective mass provides a sufficient IR screening.
  
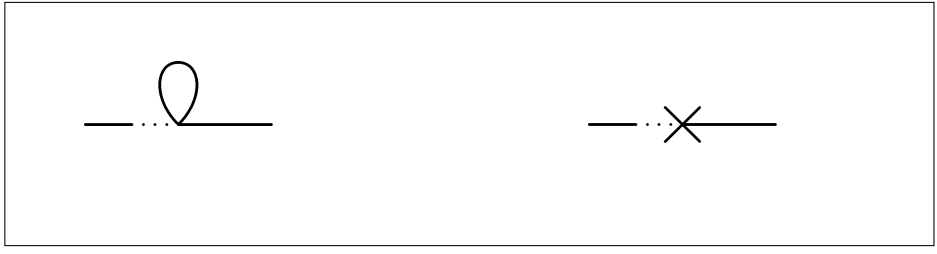
\begin{figure}[btp]
\centering
\fbox{\parbox[c][3cm]{0.8\textwidth}{
\qquad\input{tadpoleA.tex}\hfil\hfil
\input{tadpoleAc.tex}}}
\caption{Tadpole diagram and its counter-term for a quartic self interacting scalar field theory. 
The mirror diagrams should also be considered and give the same contributions. \label{fig:A}}
\end{figure}

Let us now compute the one-loop corrections to the tree-level power spectrum. 
They are given by the tadpole diagrams in Figure~\ref{fig:A} plus the same contributions given by the mirror diagrams.
In those diagrams the only change is in the external legs and the interchange $ G^R $ with a $ G^A $, so that
both give exactly the same contribution.  We will take the mirror diagrams into account including a factor 2
in the integration.

The integral corresponding to the one-loop diagram in Figure~\ref{fig:A}, with the additional
factor 2, reads 
\begin{equation}\label{eq:tadpole}
2\int_{\tau_\text{in}}^\tau d \tau_1\, \left(-iG^R(k,\tau,\tau_1)\right)F(k,\tau_1,\tau)\left[ \left(\frac{-i\lambda}{2} a^4(\tau_1)\right) \int \frac{d p^3}{(2\pi)^3} F(p,\tau_1,\tau_1)\right],
\end{equation}
where $F$ and $G^R$ are the two-point functions of our theory in the Schwinger basis. 
On a quasi de Sitter background for a massless scalar field theory they are given by
\begin{align}
F(k,\tau_1,\tau_2)&=\frac{H^2}{2k^3}\left((1+k^2 \tau_1\tau_2)\cos\,k(\tau_1-\tau_2)+k(\tau_1-\tau_2)\sin\,k(\tau_1-\tau_2) \right),\\
G^R(k,\tau_1,\tau_2)&=\frac{H^2}{k^3}\theta(\tau_1-\tau_2)\left((1+k^2 \tau_1\tau_2)\sin\,k(\tau_1-\tau_2)-k(\tau_1-\tau_2)\cos\,k(\tau_1-\tau_2) \right).
\end{align}

This one-loop integral can be solved analytically. We start evaluating the internal loop integral to identify and renormalize the divergence. 
First we introduce an arbitrary mass scale $M$ and we split the integral in two parts
\begin{align*}
\int \frac{d p^3}{(2\pi)^3} F(p,\tau_1,\tau_1) &=\left[ \int_0^M + \int_M^\infty \right] \frac{d p}{2\pi^2}\, p^2 F(p,\tau_1,\tau_1) .
\end{align*}
For the first integral we use the asymptotic expansion for the propagator in eq.~(\ref{eq:propeps})
\begin{align}
\int_0^M \frac{d p}{2\pi^2}\, p^2 F(p,\tau_1,\tau_1)  &=\frac{1}{2\pi^2}\int_0^M d p\, \frac{H^2}{2p} \left(p^2\tau_1^2\right)^\varepsilon 
\nonumber\\
&= \frac{H^2 \left(M^2 \tau_1 ^2\right)^{\varepsilon }}{8 \pi ^2 \varepsilon } \nonumber\\
&= \frac{H^2}{8 \pi ^2}\left(\frac{1}{\varepsilon } + \log \left(M^2 \tau_1 ^2\right)\right) + \mathcal{O}(\varepsilon),
\label{eq:one-loopIR}
\end{align}
where we expanded the last expression for small infrared cutoff $\varepsilon$. So we see that in the case of quasi-de-Sitter
space, the loop integral contains an IR divergence for massless fields, which is not present in Minkowski.
The second integral is instead UV divergent. We make it finite by introducing a physical cutoff $\Lambda a(\tau_1)$~\cite{Senatore:2009cf}
\begin{align}
\int_M^{\Lambda a(\tau_1)} \frac{d p}{2\pi^2}\, p^2 F(p,\tau_1,\tau_1)  &=\frac{1}{2\pi^2}\int_0^{\Lambda a(\tau_1)} d p\, \frac{H^2}{2p} \left(1+p^2\tau_1^2\right) \nonumber\\
&= \frac{H^2}{8 \pi ^2} \left(\tau_1^2 \left(\Lambda^2 a^2(\tau_1) -M^2\right)+2 \log \left(\frac{\Lambda a(\tau_1) }{M}\right)\right).
\end{align}
The total integral is finally given by
\begin{align}
\left[ \int \frac{d p^3}{(2\pi)^3} F(p,\tau_1,\tau_1)\right]_{\text{reg}} &=  \frac{H^2}{8 \pi ^2}\left(\frac{1}{\varepsilon } + \log \left(M^2 \tau_1^2\right)+\tau_1^2 \left(\Lambda^2 a^2(\tau_1) -M^2\right)+2 \log \left(\frac{\Lambda a(\tau_1) }{M}\right)\right) \nonumber\\
&=  \frac{H^2}{8 \pi ^2}\left(\frac{1}{\varepsilon } + 2 \log \left(\frac{\Lambda}{H}\right)+\left(\frac{\Lambda}{H} \right)^2\right),
\end{align}
where in the second line we have sent the mass scale $M$ to 0 since it is arbitrary and this limit is finite as
$M$ cancels in the logarithms. 
We have then the amputated amplitude as
\begin{align}
\left[ A_\text{amp} \right]_{\text{reg}} &=  - i \frac{\lambda a(\tau_1)^4 H^2}{8 \pi^2} 
\left(\frac{1}{\varepsilon } + 2 \log \left(\frac{\Lambda}{H}\right)+\left(\frac{\Lambda}{H} \right)^2\right).
\end{align}
We renormalize the amplitude in the minimal subtraction scheme, by identifying and subtracting the UV divergences from 
the original expression and by defining the time-independent counter-terms~\footnote{Note that when we insert the counter-term in 
the Lagrangian, we obtain also an additional factor $ a(\tau)^4 $, matching the corresponding factor in $ \left[ A_\text{amp} \right]_{\text{reg}} $.
Moreover we have here to add the counter-term diagrams to both of the one-loop diagrams considered.}
\begin{eqnarray}
\delta m^2 &=&  \frac{\lambda}{16\pi^2} \Lambda^2, 
\\
\delta \xi &=&  \frac{\lambda}{8\pi^2} \frac{1}{12} \log\left(\frac{\Lambda}{\mu} \right)\; ,
\label{eq:FRW-counterterms}
\end{eqnarray}
for the case $ \xi =0 $. 
We see that the amplitude has indeed the same quadratic and logarithmic divergencies as in Minkowski 
space and it is very similar, just with the identification $ m^2 = - H^2(\tau_1) $ in the prefactor of 
the logarithmic term. We can interpret this correction proportional to $ H^2 $ as a covariant 
correction of the curvature term 
$ \xi {\cal R} \phi^2 $, since in FRW we have $ {\cal R}(\tau) = 12 H^2(\tau) $.
The latter result for $ \delta\xi $ is in agreement with the results of dimensional regularization~\cite{Baacke:2010bm} 
and of the effective action method~\cite{Markkanen:2013nwa}.

So we obtain the finite amputated loop amplitude as
\begin{align}
\left[ A_\text{amp} \right]_{\text{ren}} &=  - i \frac{\lambda a(\tau_1)^4 H^2(\tau_1)}{8 \pi^2} 
\left(\frac{1}{\varepsilon } - 2 \log \left(\frac{H(\tau_1)}{\mu}\right) \right).
\end{align}
So we see that the result has only a logarithmic dependence on $ H(\tau_1) $, consistent with the result of~\cite{Senatore:2009cf},
while the factor of $ a(\tau_1)^4 $ is just related to the definition of the vertices and cancels out when we consider 
the external legs. So the amputated loop amplitude is only mildly dependent on the internal time $\tau_1$.

\begin{figure}[tbp]
	\centering\includegraphics[width=.6\textwidth]{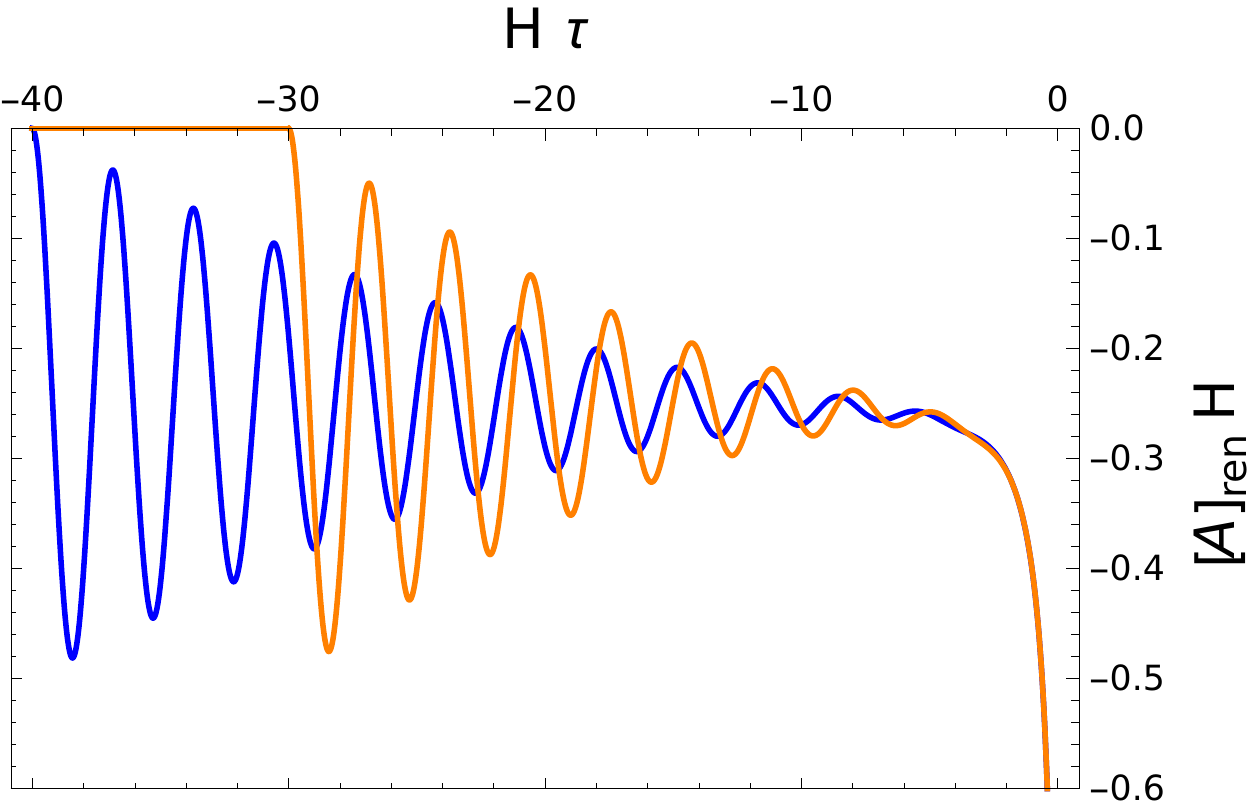}
	\caption{Time dependence of the renormalized 2-point correlation function for $H \tau_\text{in}=-40$ (blue curve) and $H \tau_\text{in}=-30$ (orange curve) with $k/H=1$, $\varepsilon=1/(16\pi^2)$, $H/\mu=1$ and 
$\lambda =1 $.}\label{fig:oscillat}
\end{figure}

Now we can compute the full correlation function including the external propagators as in eq.~(\ref{eq:tadpole}). 
We compute the integral analytically using the massless scalar field propagators in de Sitter spacetime,
which is a good approximation for all scales $k$ which are superhorizon. 
The result is given by
\begin{align}
\frac{- i }{6 k^3}
\left[ A_\text{amp}\,  a(\tau_1)^{-4} \right]_{\text{ren}} 
\times &\Bigg[ 2 +  \Big(
{\rm Ci}(2 k \tau)-{\rm Ci}(2 k \tau_{\rm in}) \Big)\Big(\left( -1 + k^2 \tau^2   \right)\cos(2k\tau)-2k\tau \sin(2k\tau)  \Big)
\nonumber\\
&+\Big( {\rm Si}(2 k \tau) - {\rm Si}(2 k \tau_\text{in}) \Big) \Big( 2k\tau \cos(2k\tau) + (-1 + k^2 \tau^2)\sin(2k \tau)  \Big) \nonumber\\
&+ \frac{1}{2 k^3 \tau_\text{in}^3}
\left(4 k^2\tau_{\rm in} \tau + \left( 1+k^2  \tau_\text{in}^2 \right) \left(1 -k^2\tau^2 \right)  \right)\sin\left(2k(\tau-\tau_\text{in})  \right) \nonumber\\
&+ \frac{1}{2 k^3 \tau_\text{in}^3}
\left( 2 k \tau_\text{in} \left( 1-  k^2 \tau^2 \right) - 2 k \tau \left( 1 + k^2 \tau_{\rm in}^2 \right)\right) \cos\left(2k(\tau-\tau_\text{in})  \right) \Bigg],
\label{eq:externaloscillation}
\end{align}
where ${\rm Ci}$ and ${\rm Si}$ are the sine and cosine integral and are defined as
\begin{align*}
{\rm Ci}(x) &= -\int_x^\infty dt\,\frac{\cos t}{t},\\
{\rm Si}(x) &=\int_0^x dt\,\frac{\sin t}{t}.
\end{align*}

 At lowest order in $k\tau$ the solution simplifies to
\begin{equation}
\frac{\lambda H^2(\tau)}{4\pi^2 12 k^3} \left[\left(\frac{1}{\varepsilon}- 2\log\left(\frac{H}{\mu}  \right)\right) \left(\log \left(\frac{\tau}{\tau_\text{in}}  \right)+\frac{1}{3}-\frac{\tau^3}{3\tau_\text{in}^3}\right)\right],
\end{equation}
where $\varepsilon$ is the infrared regulator and $\mu$ is an arbitrary energy scale used in the renormalisation procedure. 
We observe here a logarithmic dependence on the conformal time and a few polynomial terms in 
$ \frac{\tau}{\tau_\text{in}} $. Those secular terms just correspond to the first-order approximation (for small $k \tau$) 
of the early time oscillations, as can be seen from the complete solution, shown in Figure \ref{fig:oscillat} for different initial times.

One can recognize the logarithmic behaviour at late time (about $\tau=0$) and the spectral oscillations
connected to the initial state, as in the Minkowski case. In this case though, we cannot simply take the limit
of $ \tau_\text{in} \rightarrow -\infty $ as we set the Bunch-Davies vacuum at the beginning of inflation~\cite{Bunch:1978yq, Allen:1985ux}.
It is nevertheless clear that the longer the inflationary epoch, the more damped is the amplitude of the oscillations.
This is in contrast with the results obtained using the Hubble scale at the beginning of inflation as IR cut-off, since
in that case the effects becomes larger for a longer inflationary epoch~\cite{Sloth:2006az, Riotto:2008mv}.

\begin{figure}[tbp]
	\centering\includegraphics[width=.6\textwidth]{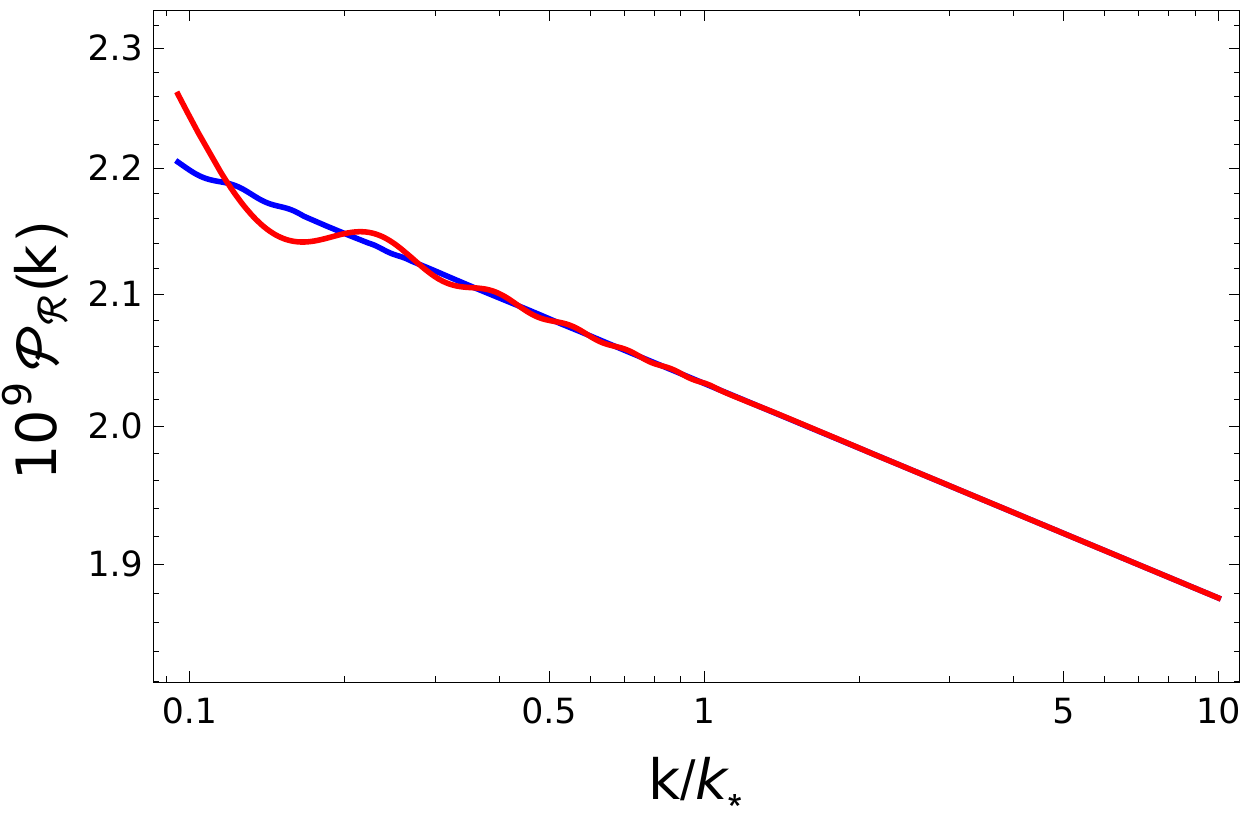}
	\caption{Power spectrum for a scalar field theory with monomial interaction with $\lambda=2.97\times10^{-13}$. The renormalization scale was set to $\mu=10^{16}\,$GeV and the initial times to $\tau_{\rm in} = -1/k_* \exp(N_{\rm tot}-N_*)$ with $N_{\rm tot} = 62$, $k_*=0.05\, {\rm Mpc}^{-1}$ and $N_*=57.5$ (blue line) or $N_*=59$ (red line). 
	The corrections to the tree-level are amplified by a factor $1.5\times10^{12}$.}\label{fig:powersplambda}
\end{figure}

Because the correlation function and the primordial spectrum are related, we expect to see the imprint of these 
early oscillations also in the power spectrum, as it is evaluated at horizon crossing. 
The tadpole diagram gives us the one-loop correction to $\mathcal{P}_{\mathcal{R}}$ using the perturbative expansion in 
eq.~(\ref{eq:PowerSpectrum}).  
The correction is proportional to $\lambda$ which is fixed to be very small from the normalization in eq.~(\ref{eq:lambda}). 
In Figure \ref{fig:powersplambda} we show our results for the first-order correction to the power spectrum for different 
initial times where we amplified the corrections by a factor $1.5\times10^{12}$. Here we take the IR-cutoff of order of the slow-roll 
parameters $ 2\varepsilon \sim 2\eta = 3 \epsilon $. Since  for the case of a quartic potential the slow-roll parameters are
of order ${\cal O}(0.1) $, the enhancement from the IR divergence is in this case limited.
Note also that the oscillatory behaviour is different compared to other models with features, see eg. \cite{Chluba:2015bqa} for
a review of the different models.
For a long inflationary phase, when the oscillations are damped, the one-loop correction gives a tiny constant shift of the 
power spectrum, which can be reabsorbed into a redefinition of the coupling $ \lambda $ corresponding to the overall spectrum 
normalization in eq.~(\ref{eq:lambda}). This shift is visible in Figure \ref{fig:powersplambda} only because of the large
amplification factor used.

As one can see from the figure, the one-loop corrections are very small because their order of magnitude is fixed 
by the coupling constant $\lambda$. We found a departure from a near-scale invariant power spectrum of 
\begin{equation}
\left|\frac{P_{\cal R}^{\rm loop} (k) }{P_{\cal R}^{\rm tree} (k) }\right|_\text{quartic}\leq  0.5 \times 10^{-13}.
\end{equation}

The correction is compatible with the current observations and is too small to be observed in future experiments. \\

\section{One-loop corrections in quasi de Sitter space from a massive field}

To have a richer phenomenology we investigate also the hybrid inflation model that describes an inflaton field $\phi$ 
and a hybrid field $\sigma$ with the effective potential (\ref{eq:hybrid}). In this model one has to consider the loop 
corrections  from the massive scalar field during inflation.

We compute the amputated tadpole using both the WKB and the full propagator given by the Fourier transform of
the hypergeometric function. The integration is in both cases dominated by the UV region and there the scale factor
suppression of the propagator $ \propto a^{-2} (\tau) $ is compensated by the choice of a physical UV cut-off 
$ \Lambda a(\tau) $ in momentum space. Therefore we do not find any suppression by negative powers of 
$ a(\tau) $, contrary to what happens for the IR modes~\cite{Weinberg:2006ac}.

 In the WKB case we obtain after renormalization the analytic expression
\begin{equation}
[A_{\rm amp}]_{\rm ren}= \frac{-i\lambda_h^2 m^2 a^4(\tau)}{4\pi^2}\left(1+\log\left(\frac{m^2}{4\mu^2} \right) \right).
\label{eq:Aamp-WKB}
\end{equation}
The full result allows us to extract the UV divergence and define an appropriate mass counter-term.
It is interesting to note that the it contains also a term that is not present in the massless case, 
i.e. the logarithmic divergence proportional to the squared mass:
\begin{equation}
\delta m^2=  \frac{\lambda_h^2}{4\pi^2}\left(\Lambda^2 - m^2\log\left(\frac{\Lambda}{\mu} \right) \right). \; 
\label{eq:deS-counterterm}
\end{equation}
Therefore, it is not sufficient to renormalize the tadpole computed with the full massive propagator by 
substracting only the UV divergences of the massless case, as a residual logarithmic divergence survives.  
Our strategy is to subtract instead the counter-term computed analytically with the WKB propagator here
above, identical to the Minkowski one in eq.~(\ref{eq:Mink-counterterm}), as well as the counter-term for the 
curvature term in eq.~(\ref{eq:FRW-counterterms}), which does not appear in the WKB approximation.

\begin{figure}[tbp]
	\centering\includegraphics[width=.6\textwidth]{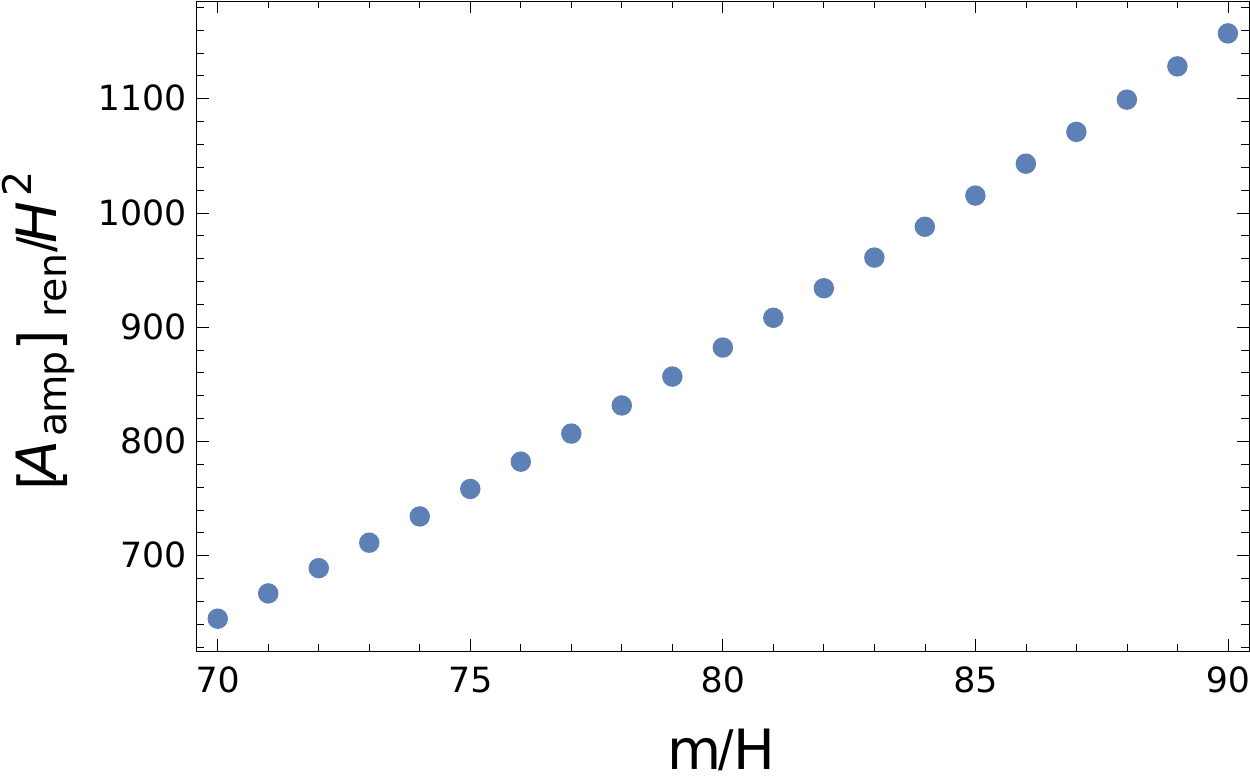}
	\caption{Mass dependence of the amputated tadpole diagram after renormalization with the WKB couter-term.}
	\label{mass}
	\end{figure}

In Figure~\ref{mass}  we show the $m^2$ dependence of the finite part of the amputated diagram, which 
matches well that predicted by the approximated expression in eq.~(\ref{eq:Aamp-WKB}). 
Indeed the contribution grows quadratically with the mass of the field, so that the heavier fields give the strongest effect.
So while in the case of  (nearly) massless fields, the correction was proportional to the Hubble
parameter $ H^2(\tau) $, now it is proportional to $ m^2 $, larger than $ H^2(\tau) $ if the
field is not dynamical during inflation.

We also estimated analytically the best parameter choice for models given by eq.~(\ref{eq:hybrid}) using the
WKB approximation, which gives a quite good fit to the full result. 
In this case the mass in the amputated amplitude eq.~(\ref{eq:Aamp-WKB}) is the mass of the heavy field 
$m_\sigma^2 = 2 \lambda_h^2 \phi^2 - M^2$. If the inflaton dynamics is dominated by the mass term,
then one has from eq.~(\ref{eq:Efolds})
\begin{equation}
m_\sigma^2 \simeq 8 \lambda_h^2 M_P^2 N(k),
\end{equation}
where $N(k)$ is the number of e-folds from the time the scale $k$ exits the horizon to the end of inflation. 
We see that the mass of the hybrid field is slowly varying during inflation and the change is always adiabatic 
for small $k$, since
\begin{equation}
\frac{\dot w_k}{w_k^2} \sim \frac{\dot m_\sigma}{m_\sigma^2} = \frac{H}{2\; m_\sigma\; N} \ll 1.
\end{equation}
We will therefore use this slowly varying mass in the massive propagator.

The renormalized WKB amplitude (\ref{eq:Aamp-WKB}) becomes
\begin{equation}
[A_{\rm amp}]_{\rm ren}= \frac{-2 i \lambda_h^4}{\pi^2} \frac{N(k) M_P^2}{H^4 \tau^4} 
\left(1+\log\left( \frac{\lambda_h^2 M_P^2 N(k)}{\mu^2} \right) \right).
\end{equation}
The one-loop correction to the power spectrum is then given by
\begin{equation}
P_{\cal R}^{\rm loop} (k) \approx \frac{k^3}{4\pi^2}\left( \frac{\lambda_h^4}{\pi^2}\frac{N(k) }{\epsilon}\right)(-i f(k, \tau, \tau_\text{in})),
\end{equation}
where $f$ is an oscillatory function that includes the contribution coming from the external propagators, as given in 
eq.~(\ref{eq:externaloscillation}).
The loop correction has to be compared to the dominant contribution to the power spectrum given by the tree-level
\begin{equation}\label{eq:Ptree}
P_{\cal R}^{\rm tree} (k) = \frac{H^2}{8 \pi^2 \epsilon M_P^2},
\end{equation}
giving a correction of the order
\begin{equation}
\frac{P_{\cal R}^{\rm loop} (k) }{P_{\cal R}^{\rm tree} (k) } = \frac{2 \lambda_h^4 N(k)}{ \pi^2}\,\frac{M_P^2}{H^2}\,(-i k^3 f(k, \tau, \tau_\text{in})),
\end{equation}
where $\left|-i k^3 f(k, \tau, \tau_\text{in})\right| = C(k) $. For the case of a very long inflation, when the oscillations are negligible,
such factor $ |C (k) | $ reduces to a constant of order $ \sim 0.2 $. 

To have a correction of the order of $10^{-2}$ we need a coupling constant of the order of
\begin{equation}
 \lambda_h^2 \sim \frac{\pi}{10} \frac{H}{M_P \sqrt{2N(k) C(k)}}.
\end{equation}
Moreover, the Hubble constant can be estimated from (\ref{eq:Efolds})
\begin{equation}\label{eq:Hinf}
H^2 = \frac{1}{3M_P^2} \frac{m^2\phi^2}{2} = \frac{2}{3} m^2  N(k)
\end{equation}
and the coupling constant has to be small enough to remain in the case of mass-dominated inflation, i.e.
from eq.(\ref{eq:lambdahsmall}),
\begin{equation}
 \lambda_h^2 < \frac{\pi}{\sqrt{N(k)}} \frac{m}{M_P}\; ,
\end{equation}
 so the maximal loop correction that we can obtain is
\begin{equation}
\left|\frac{P_{\cal R}^{\rm loop} (k) }{P_{\cal R}^{\rm tree} (k) }\right| < \frac{3\;C(k)}{N(k)}\, \sim 10^{-2}.
\end{equation}
So we see that in this case no large corrections can be reached, as the coupling is bounded from above and
we have a suppression by $ 1/N(k) $ and by the oscillatory factor $ C (k)$.

Let us now investigate instead the case of large coupling $ \lambda_h^2 $. In particular we would like to study the 
radiative corrections to the power spectrum for a hybrid model inspired by a supersymmetric theory with the 
scalar potential given in eq.~(\ref{eq:suppot}), when the coupling also determines the slope of the inflaton potential
due to the one-loop Coleman-Weinberg potential. In that case,
as discussed earlier, an inflationary phase is possible for $\lambda_h $ up to the order of $ 10^{-3}$.
But, as in the non-supersymmetric case, here the mass of the hybrid field is determined by the inflaton V.E.V.
and therefore the effect cannot be arbitrarily large.
\begin{figure}[tbp]
	\centering\includegraphics[width=.6\textwidth]{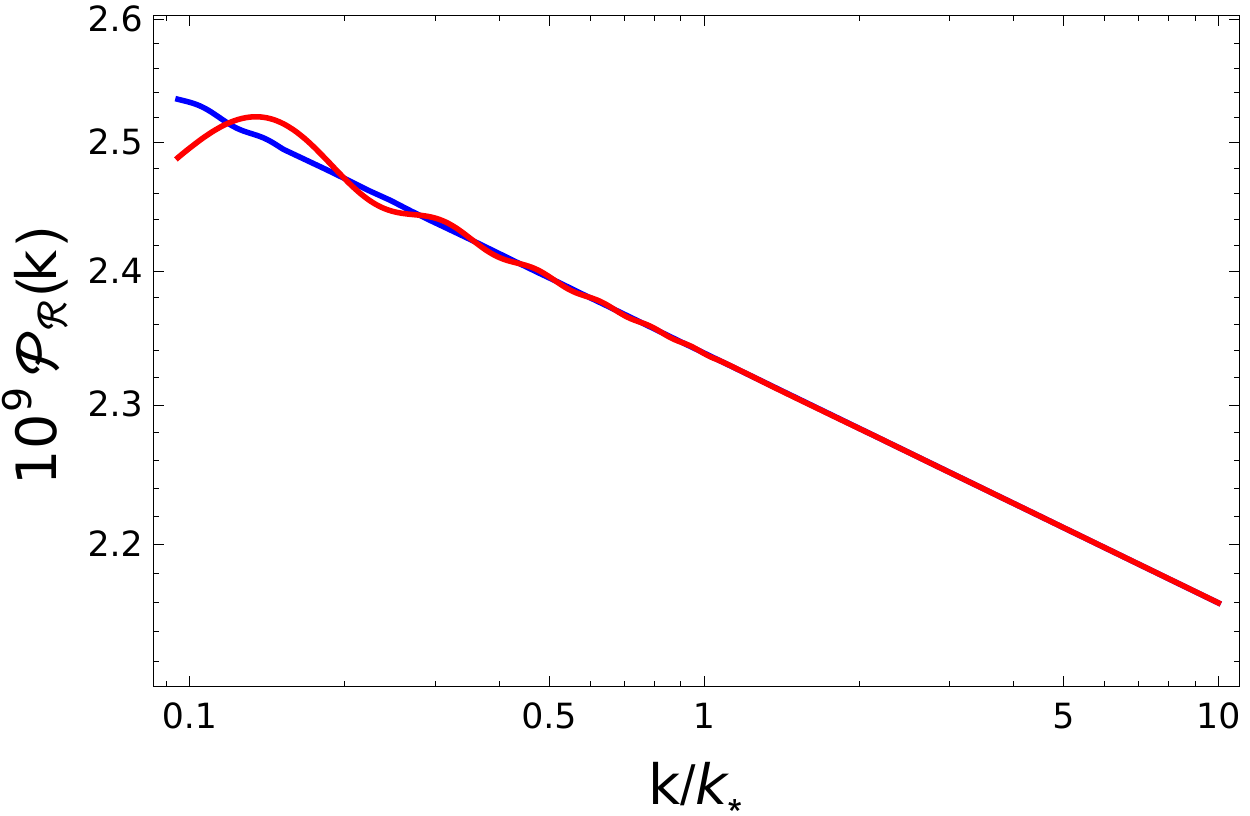}
	\caption{Renormalized power spectrum for hybrid inflation obtained with the full propagator with 
	$\lambda_{h} = 3\times10^{-3}$ and $\mu$ of the order of $3\lambda_h M_G$. The initial times are set to $\tau_{\rm in} = -1/k_* \exp(N_{\rm tot}-N_*)$ with $N_{\rm tot} = 62$, $k_*=0.05\, {\rm Mpc}^{-1}$ and $N_*=57.5$ (blue line) or $N_*=59$ (red line). 
	\label{fig:hybridspectrum} }
\end{figure}

From  (\ref{eq:suppot}) we can estimate the order of the mass of the scalar component of the superfield $\Sigma$ and
the inflationary scale $ H$ as:
\begin{align*}
m_\Sigma^2 &= 2\lambda_h^2 (\phi^2  -M_G^2 ) , \quad\quad
3H^2 \approx \frac{V}{M_P^2}=\lambda_h^2 \frac{M_G^4}{M_P^2}.
\end{align*}
Therefore we obtain
\begin{align}
\frac{m_\Sigma^2}{H^2}\approx  \frac{\lambda_h^4 M_P^2 N}{\pi^2}\; \frac{3 M_P^2}{\lambda_h^2 M_G^4} =
 \frac{3 \lambda_h^2 N\; M_P^4}{\pi^2\; M_G^4} \sim 1.35 \times 10^7\; ,
\end{align}
so we see that the one-loop contribution is in this case enhanced with respect to the  quartic scalar field inflation 
by a factor of the order $ 10^7 $ due to the hybrid field's mass, which grows towards the beginning of inflation.

The power spectrum is shown in Figure \ref{fig:hybridspectrum} and was computed numerically with the full propagators 
with the renormalization prescription described above and setting the initial conditions at the beginning of inflation. 
The oscillations in the power spectrum emerge when the external propagators are included. 
The increase in the effect is mainly due to the fact that the corrections to the power spectrum are proportional to the coupling 
$\lambda_{h}^2 \gg \lambda $, but also the heaviness of the hybrid field enhances the size of the corrections substantially. 
Unfortunately the sensitivity of the instruments today is not enough to detect an effect that in the most optimistic 
case gave us a departure of a nearly-scale invariant spectrum of
\begin{equation}
\left|\frac{P_{\cal R}^{\rm loop} (k) }{P_{\cal R}^{\rm tree} (k) } \right|_\text{hybrid}\leq 0.7 \times 10^{-1}.
\end{equation}

Generically we can obtain an analytical estimate of the largest possible effect, while still retaining the power spectrum normalization
in accordance to the CMB data and slow-roll and this is similar to the bound obtained before for the case of mass-dominated
dynamics, but not quite equal. We have in this case from the maximal value of the coupling in eq.~(\ref{eq:lambda_h_max}) and
the value of the Hubble parameter during inflation:
\begin{equation}
\left|\frac{P_{\cal R}^{\rm loop} (k) }{P_{\cal R}^{\rm tree} (k) } \right|< 0.12\, N(k) C(k) \, \sim 1.
\end{equation}
So in this case we can indeed reach larger values of the correction. 
Note though that here we are only considering the one-loop contribution of the hybrid scalar field, 
while in supersymmetric models also the corresponding fermionic fields are 
present and would give a negative contribution, which will partially cancel the effect~\footnote{We expect indeed the fermionic 
contribution to cancel exactly the quadratic divergence in the loop, but still a contribution proportional to the mass difference 
between the scalar and fermion masses should survive as supersymmetry is broken spontaneously and explicitly, as
given by the $ M_S^4$ term, during inflation, see \cite{Buchmuller:2000zm}.}.

To conclude our analysis we analytically estimated the correction to the power spectrum for the case of an heavy spectator 
field. The renormalized amplitude is given by (\ref{eq:Aamp-WKB}) where $m$ is the spectator's mass $m_\sigma$. 
The one-loop correction to the power spectrum is therefore given by

\begin{equation}
P_{\cal R}^{\rm loop} (k) \approx \frac{k^3}{4\pi^2}\left(\frac{\lambda_h^2 m_\sigma^2}{8\pi^2 \epsilon M_P^2}  \right)
\left(1+\log\left( \frac{m^2_\sigma}{4\mu^2} \right) \right)(-i f(k, \tau, \tau_\text{in})).
\end{equation}
In order to obtain the largest effect we consider the maximum value for the spectator's mass $ m_\sigma \sim 10^{-2} M_P $
and the coupling $\lambda_h^2 \sim 10^{-6} $ and we found a maximal correction of

\begin{equation}
\left|\frac{P_{\cal R}^{\rm loop} (k) }{P_{\cal R}^{\rm tree} (k) } \right|=
\frac{3 \lambda_h^2}{8\pi^2} \frac{m_\sigma^2}{m^2} \frac{C(k)}{ N(k) }
 <  0.4 \times 10^{-3}, 
\end{equation}
where $P_{\cal R}^{\rm tree}$ is given in (\ref{eq:Ptree}) and the Hubble constant was estimated from (\ref{eq:Hinf}). 
We see here that the expression resembles the one for the chaotic inflationary case, but it is enhanced by the larger mass
and coupling. Nevertheless, the effect of the radiative correction for a quadratic inflation model with one spectator field,
which does not affect the background evolution, is smaller than in the case of the hybrid models that we discussed earlier,
since it is suppressed by the number of e-folds.

Note that if the spectator field is even heavier, its contribution to the energy density of the Universe will take over in
the inflaton potential and then the correction will be suppressed by negative powers of $m_\sigma^2 $, so that
the heavy field will effectively decouple.

\section{The 4-point correlation function}

After considering the correction to the 2-point function, we now proceed to consider the next order, the 4-point function,
which gives a contribution to the trispectrum~\cite{Seery:2006vu, Byrnes:2006vq}.
Indeed the 3-point function $\left< \delta\phi_{k_1}\delta\phi_{k_2} \delta \phi_{k_3}\right>$ and all the
other odd correlation functions are identically zero at the first order in cosmological perturbation theory 
in the models we investigated because the Lagrangian of the form 
eq.~(\ref{eq:lagrang}) is invariant under the field transformation $\phi \to -\phi$. 
Higher order correlation functions of the inflaton fluctuations are additional quantities that can be theoretically predicted 
from inflationary models, give information about the non-Gaussian nature of the primordial field(s)
and can also be used  for disentangling among different inflationary models.
The last Planck data release \cite{Ade:2015ava} yielded new constraints on primordial 
non-Gaussianities and they found no evidence for such features.

We will now apply the CTP formalism and investigate the time-dependence of the radiative corrections to higher orders correlation 
functions with the same strategy used for the two-point function. We will first study the two tree-level contributions with a time 
dependence coming from the vertex $a^4(\tau)\lambda$ and give our prediction for the cosmological parameter 
$\tau_{\rm NL}$. Subsequently we will estimate the first-order radiative corrections. 
At one-loop there is only the fish diagrams contributing, giving corrections proportional to $\lambda^2$.

\subsection{Tree-level contributions}
At tree-level we have two finite diagrams $T_1$ and $T_2$ contributing to the 4-point correlation function. 
The Feynman rules give us the following expressions 
\begin{table}[h!]
\begin{tabular}{cc}
\input{tree1.tex} & \raisebox{6mm}{\small{$= 6 (-i)\left(\frac{-i\lambda}{3!}\right) \int_{\tau_\text{in}}^\infty  d\tau_1\,a^4(\tau_1) G^R(k_1, \tau, \tau_1)F(k_2, \tau, \tau_1)F(k_3, \tau_1, \tau)F(k_4, \tau_1, \tau),$}}\\[5mm]
\input{tree2.tex}&\raisebox{6mm}{\small{$= 6i\left(\frac{-i\lambda}{4!} \right)\int_{\tau_\text{in}}^\infty d\tau_1\, a^4(\tau_1) G^R(k_1, \tau,\tau_1)F(k_2,  \tau,\tau_1)G^A(k_3, \tau_1, \tau)G^A(k_4, \tau_1, \tau),$}}
\end{tabular}
\end{table}
where $k_1, k_2, k_3$ and $k_4$ are the momenta and $\tau$ is the conformal time of the external propagators. 

The integrals are computed analytically using the super-Hubble approximation $|k \tau_i|\ll 1$ for the massless external propagators $F$ and $G^R$. 
By considering all possible permutations of the external momenta we obtain the explicit expressions:
\begin{align}\label{eq:T1T2}
T_1&=- \lambda \frac{H^4 \left(k_1^3+k_2^3+k_3^3+k_4^3 \right)\left( \frac{ \tau ^3}{\tau_\text{in}^3}+3  \log \left(\frac{\tau_\text{in}}{\tau }\right)-1\right)}{72 k_1^3\,k_2^3\, k_3^3\, k_4^3}, \\
T_2&= \lambda \frac{H^4 \left( k_2^3k_3^3k_4^3+k_3^3k_4^3k_1^3+k_4^3k_1^3k_2^3+k_1^3k_2^3k_3^3\right)\left(  2 \frac{\tau^9}{ \tau_\text{in}^3}+3 \tau ^6-6 \tau ^3\tau_\text{in}^3+\tau_\text{in}^6+18 \tau ^6 \log \left(\frac{\tau_\text{in}}{\tau }\right)\right)}{1296 k_1^3\,k_2^3\,k_3^3\,k_4^3} \nonumber. 
\end{align}

\begin{figure}[tbp]
\centering
\includegraphics[width=.6\textwidth]{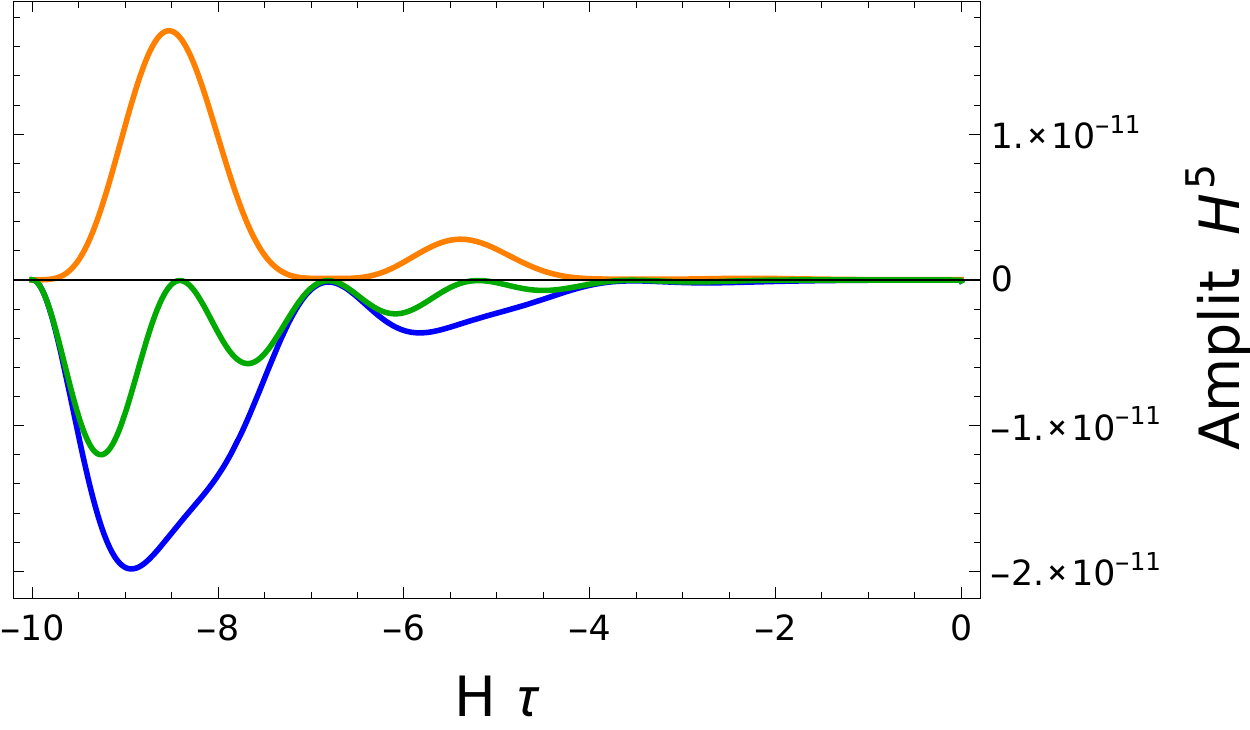}
\caption{Tree-level contributions $T_1$ (blue line), $T_2$ (yellow line) and $T_1+T_2$ (green line) using the full expression for the propagators for $k_{1,2,3,4}/H=1$, $\lambda=10^{-13}$ and $H\tau_\text{in}=-10$.}\label{fig:tree}
\end{figure}

We observe a similar logarithmic dependence on the conformal time as we had for the two-point function. 
The secular terms in $\frac{\tau}{\tau_{\rm in}}$ are the first-orders of the early time oscillations that one can see in Figure \ref{fig:tree},
where we give the full result with no expansion in $|k \tau_i |\ll 1$.

In Figure \ref{fig:tree}, one can recognize the spectral oscillations connected to the initial state. The logarithmic dependence $\log(\tau)$ of $T_1$ is present but it is not visible in the plot. The oscillations are very small and the amplitude is fixed by the normalization of $\lambda$ in eq~(\ref{eq:lambda}).

\subsection{One-loop contributions}
We consider now the first-order corrections to the 4-point function. This computation involves the product of two singular propagators 
and has to be renormalized. For example one of the divergent integral is of the following form, where we omitted the external propagators,
\begin{equation}
36 (-i) \left(\frac{-i \lambda}{3!}\right)^2\int \frac{d^3 \vec{p}_1}{(2\pi)^3}\int \frac{d^3 \vec{p}_2}{(2\pi)^3} F(p_1, t_1, t_2)G^R(p_2,t_1,t_2)(2\pi)^3\delta(k_1+k_2-p_1-p_2).
\end{equation}
These integrals are the same as those that appear in the one-loop corrections to the two-point function for a $\lambda \phi^3$-scalar field theory 
and were already computed in \cite{vanderMeulen:2007ah} for the case of a massless field in de Sitter. We will use their results in the following. Indeed, the integration in the momentum variables cannot be performed analytically without any simplification of the propagators. 
In the following calculations we will assume that the external momenta are super-Hubble, i.e. with wavelengths above the Hubble radius
and consider the virtual particle to be massless, apart in the IR region where we use $ \nu = 3/2 - \varepsilon $ as an IR regulator.

Let us consider the finite contribution of the 4-point function of diagrams $B$, $C_1$ and $C_2$. We can see that in this case the product of distributions $F^2$, $(G^R)^2$ and $(G^A)^2$ diverge linearly, but the sum $(G^A)^2 + (G^R)^2 - F^2$ is finite.
We give the details of the computation in Appendix~\ref{ap:4pointBC}.
From the sum of the diagrams $B$, $C_1$ and $C_2$, we computed the analytic expression
\begin{multline}
\frac{H^4 \lambda ^2 \left(k_1^3+k_2^3\right) \left(k_3^3+k_4^3\right) \left(3 \log \left(\frac{\tau}{\tau_\text{in}}\right)+1\right)}{15552 \pi ^2 (k_1+k_2)^3 k_1^3 k_2^3 k_3^3 k_4^3  } \Big(9 \log \left(\frac{\tau}{\tau_\text{in}}\right) \left(2 \log \left(k^2 \tau \,\tau_\text{in}\right)+\frac{1}{\varepsilon}\right)\\
+12   \log (-k \tau)+4  +\frac{3}{\varepsilon}\Big).
\end{multline}
As for the 2-point function the correction shows a logarithmic dependence on the conformal time, in this case double, and 
the first-order contributions of the series expansion of the oscillatory terms. It also is IR divergent as expected.

Similarly we study the contribution of diagrams $A_i$~s, that involve the product of the propagators $F$ and $G^{A/R}$. 
In this case the integrals are UV divergent and we need to renormalize such divergence with the introduction of the counter-term
\begin{equation}
\delta \lambda = - 3\frac{\lambda^2}{16 \pi^2} \log{\frac{\Lambda}{\mu}},
\end{equation}
which fully agrees with the Minwkowski counter-term \cite{Ramond:1981pw} and the computations in different schemes in
 \cite{Baacke:2010bm} and \cite{Markkanen:2013nwa}.
The analytic expressions of the corrections including the external propagators in the limit $ | k \tau | \ll 1 $ are given in 
Appendix~\ref{ap:4pointA}.

\begin{figure}[tbp]
\centering
\includegraphics[width=.6\textwidth]{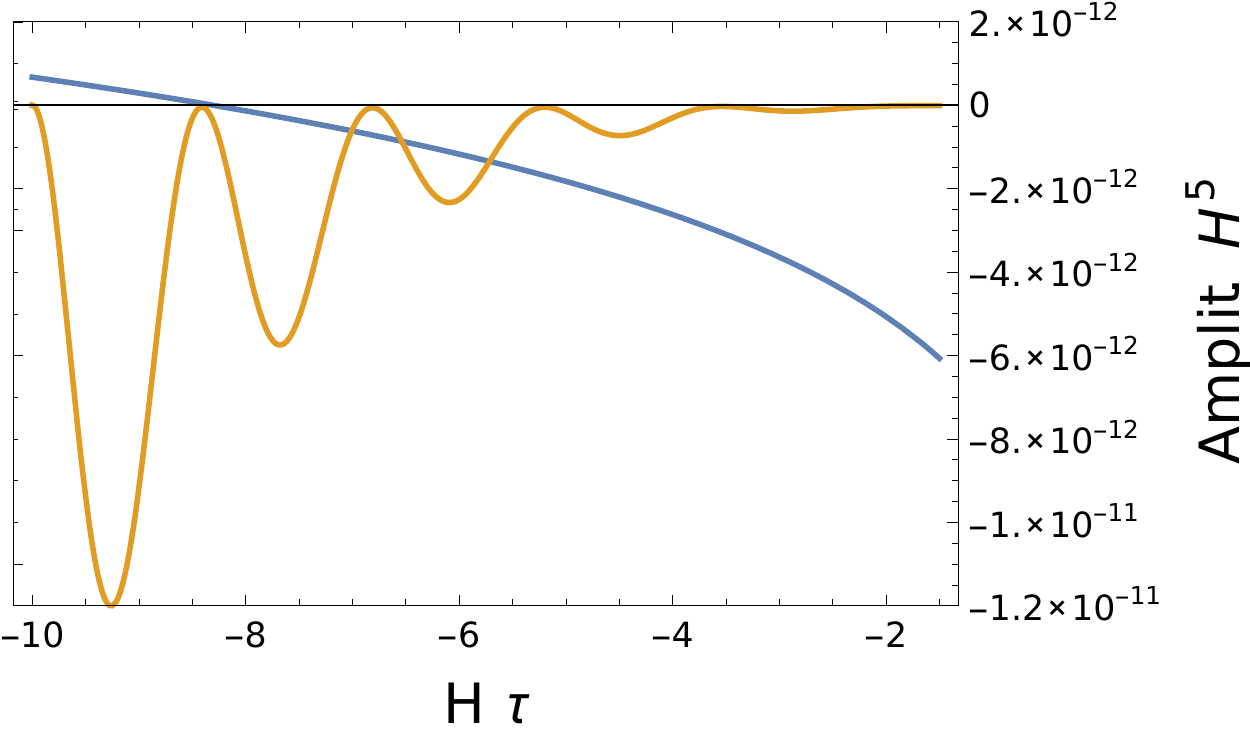}
\caption{Tree-level (yellow line) and one-loop contribution (blue line) for the 4-point function 
where we set $k_{1,2,3,4}/H=1$, $\lambda=10^{-13}$ and $H\tau_{\rm in}=-10$. 
The loop contribution is amplified by a factor $10^{13}$.}\label{fig:loop}
\end{figure}

Unfortunately, the corrections are suppressed by the overall factor $\lambda^2$ and therefore in this case the loop contributions 
are negligible compared to the tree-level result, which already contains a non-trivial time dependence.
We show the sum of all one-loop contributions in Figure~\ref{fig:loop}, where it has been enhanced by a factor of 
$ 10^{13} $. Note that the radiative correction does not show any oscillations since we have kept only the first-order
expansion for $ | k \tau | \ll 1 $, but we expect oscillations to be present in the full result. In any case the size of the
one-loop correction is so small, that we do not need to add the full external propagators as the whole one-loop contribution 
can be safely neglected.

So we see that the trispectrum,  in the case of chaotic inflation, has a non-trivial contribution already at the tree-level, 
suppressed by the coupling constant $ \lambda $, so that this effect is way beyond any possible observation.

\subsection{Non-linearity parameter}

We conclude our discussion with the calculation of the non-linearity parameter $\tau_{\rm NL}$ that can be constrained from CMB 
measurements~\cite{Ade:2015ava}. The connected trispectrum $\mathcal{T}_\mathcal{R} $ is defined as \cite{Seery:2006vu, Byrnes:2006vq}

\begin{equation}\label{eq:trisp1}
\left< \delta\phi_{\vec{k}_1}\delta\phi_{\vec{k}_2}\delta \phi_{\vec{k}_3}\delta\phi_{\vec{k}_4}\right>_c  
= \frac{\dot{\phi}^4}{H^4} (2\pi)^3 \delta\left(\sum_i \vec{k}_i  \right) 
\mathcal{T}_\mathcal{R}  (\vec{k}_1,\vec{k}_2, \vec{k}_3, \vec{k}_4).
\end{equation}
and is the first non-zero higher order correlation function for a $\lambda\phi^4$ model. A model with precise Gaussianity predicts 
$\mathcal{T}_\mathcal{R}=0$ at tree-level. The departure from Gaussianity can be described by introducing the dimensionless 
cosmological parameter $\tau_{\rm NL}$~\cite{Seery:2006vu} 
\begin{equation}\label{eq:trisp2}
\mathcal{T}_\mathcal{R} (\vec{k}_1,\vec{k}_2, \vec{k}_3, \vec{k}_4) = 
\frac{1}{2} \tau_{\rm NL}\left[\left(\frac{2\pi^2}{k_1^3}  \right) \mathcal{P}_\mathcal{R}(k_1)
\left(\frac{2\pi^2}{k_2^3}  \right)  \mathcal{P}_\mathcal{R}(k_2) \left(\frac{2\pi^2}{k_{14}^3}  \right)  \mathcal{P}_\mathcal{R}(k_{14})+ 23\; \text{permutations}  \right],
\end{equation}
where $\vec{k}_{ij}=\vec{k}_i+\vec{k}_j$ and $\sum_i \vec{k}_i = 0$.

The non-linearity parameter $\tau_{\rm NL}$ depends on the inflationary model and can be used as a test for disentangling
them. From eqs. (\ref{eq:trisp1}) and (\ref{eq:trisp2}) we get
\begin{equation}
\tau_{\rm NL} = \frac{1}{2\, \epsilon^2\, M_P^4} \frac{1}{(2\pi^2)^3} \frac{ \left< \delta\phi_{k_1}\delta\phi_{k_2}\delta \phi_{k_3}\delta\phi_{k_4}\right>_c'}{\frac{\mathcal{P}_\mathcal{R}(k_1)\mathcal{P}_\mathcal{R}(k_2)
\mathcal{P}_\mathcal{R}(k_{14})}{k_1^3 k_2^3 k_{14}^3} + 23\;\text{permutations}}\; ,
\end{equation}
where the 4-point function is evaluated when all the momenta $ k_i $ are superhorizon, i.e. for $ \min(k_i) \tau = - 1 $ and
the prime means that the  overall $\delta $-function in the momenta is factorized out.

 In the following we will consider the effects on $\tau_{\rm NL}$ coming from the initial state for the tree-level contribution $T_1$+$T_2$ of eq. (\ref{eq:T1T2}) for an equilateral configuration. In that case the expression simplifies to
 \begin{equation}
\tau_{\rm NL} (k) = \frac{1}{16\, \epsilon^2\, M_P^4\; A_s^3} \frac{k^9}{(2\pi^2)^3}  \left(\frac{k}{k_*} \right)^{3(1-n_s)} 
\frac{\left< \delta\phi_{k}\delta\phi_{k}\delta \phi_{k}\delta\phi_{k}\right>_c'}{\sum_{i=2}^4 \left[ 2(1+\cos\theta_{1i})\right]^{-3/2 + (n_s-1)/2}} \; ,
\end{equation}
 where we use the power-law expression for the power spectrum as $ \mathcal{P}_\mathcal{R}(k) = A_s \left(\frac{k}{k_*} \right)^{n_s-1} $
 and denoted with $ \theta_{1i} $ the angle between $ \vec{k}_1 $ and $ \vec{k}_i $. Note that since $ A_s $ is of order $ 1/\epsilon $
 in the slow-roll approximation, the trispectrum is at leading order proportional to $ \epsilon $. We will neglect here higher order terms
 in the cosmological perturbative expansion, suppressed by higher powers of $ \epsilon, \eta $.
 
\begin{figure}[tbp]
\centering
\includegraphics[width=.6\textwidth]{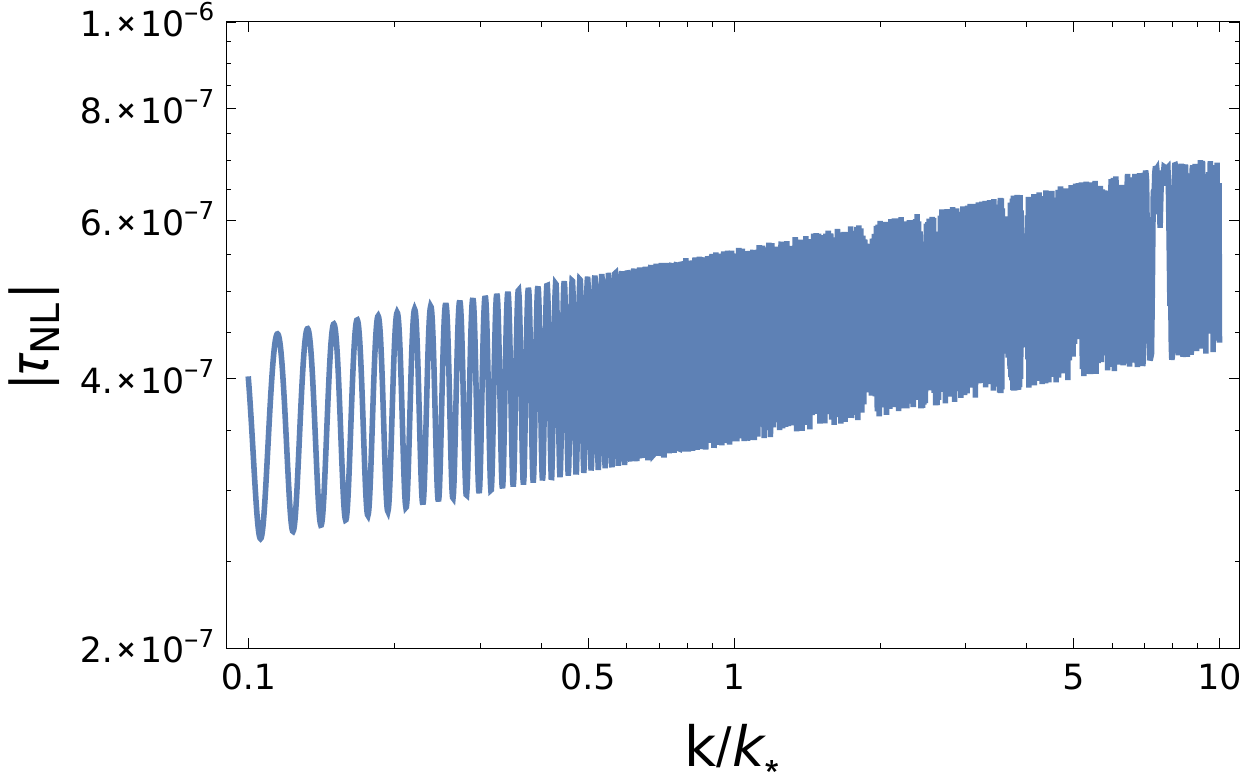}
\caption{Non-linearity parameter $|\tau_{\rm NL}|$ calculated from the contribution $T_1 + T_2$ for a 
$\lambda \phi^4$-theory with $\lambda = 2.97 \times 10^{-13}$. The initial time is set to 
$\tau_{\rm in} = -1/k_* \exp(N_{\rm tot}-N_*)$ with $N_{\rm tot} = 62$ and $N_*=57.5$.}\label{fig:tau_NL}
\end{figure}

  The value of $ \tau_{\rm NL} (k) $ is plotted in Figure~\ref{fig:tau_NL} for the maximal case of $ \cos\theta_{1i} = -1/3 $ for all $i$. 
  It is very strongly oscillating around a value of $4\times10^{-7} $ and therefore is very far from the present Planck observation 
  $ g_{\rm NL}^{local} = (-9\pm 7.7) \times 10^{4} $,  where we have $  \tau_{\rm NL} \sim \frac{18}{25} g_{\rm NL}^{local} $. 
  We see therefore that the value of the non-linearity parameter for chaotic inflation is way smaller than the present bounds.

\subsection{Massive One-Loop}

Before concluding, let us discuss briefly the case of a massive loop.
Indeed in the case of hybrid inflation instead, there is no tree-level contribution due to the absence of the $ \lambda \phi^4 $
coupling in the Lagrangian (note that such coupling is not forbidden in the simple model we consider, we just assume that it is 
negligible and smaller than in chaotic inflation), and so the one loop contribution from a loop of massive fields dominates the 
4-point function.
From the comparison of the 2-point functions one-loop corrections between the massless and massive case, we expect
the loop correction in this case to contain terms proportional to $ m_\Sigma^4, m_\Sigma^2 H^2 , H^2 $.
So the contribution will be maximally enhanced by the factor
\begin{equation}
\frac{[A_{4, \rm{amp}}]^{\rm hybrid}}{[A_{4,\rm{amp}}]^{\rm chaotic}} \propto  \frac{\lambda_h^4 m_\Sigma^4}{\lambda H^4} =  7.1 \times 10^{8}\; ,
\end{equation}
so in this case $ \tau_{\rm NL} $ may become of order 1, but still not large enough to be observed. 

For a spectator field not influencing the inflationary classical dynamics, with $ m_\sigma \sim 10^{-2} M_P$ and 
the coupling $ \lambda_h^2 \sim 10^{-6} $, we have instead
\begin{equation}
\frac{[A_{4, \rm{amp}}]^{\rm hybrid}}{[A_{4,\rm{amp}}]^{\rm chaotic}} \propto  \frac{\lambda_h^4 m_\sigma^4}{\lambda H^4}  = 2.3 \times  10^{11}\; ,
\end{equation}
which may  as well reach an interesting range, in absence of cancellations in the loop corrections.
We leave the precise computation of this contribution to future work.

\section{Conclusion}

We have investigated the one-loop radiative corrections to the primordial power spectrum for different inflationary 
models in a FRW spacetime using the Closed-Time-Path formalism. Even if the Poincar\'e symmetry in curved spacetimes 
is broken, the renormalization counter-terms can be chosen constant, and exactly as in Minkowski space, in the 
Minimal Subtraction Scheme. Their expressions coincide with those found earlier in \cite{Baacke:2010bm} with
dimensional regularization and  also with results within other formalisms, as renormalization via the effective 
action~\cite{Markkanen:2013nwa} or axiomatic field theory \cite{Hack:2015zwa, Gere:2015qsa}.
On the other hand, the finite part of the loop corrections  contains a finite logarithmic dependence on the 
slowly varying Hubble parameter and is therefore time-dependent. 
Such correction is  suppressed by the quartic coupling and strongly enhanced by the IR regulator, which we 
take here conservatively as the perturbation's effective mass $ \propto | n_s -1 | $.
Moreover  an oscillatory behaviour arises connected to the external legs of the diagram and dependent on the
initial time and (in principle) the initial conditions. Those oscillations are in general damped during the
inflationary epoch and therefore affect more strongly the large scales, that leave the horizon at the early 
stages of inflation. In the power spectrum and higher order correlations functions, this is translated to 
an oscillatory pattern in the momentum, {\it  which is intrinsic of any interacting inflaton model} even for
a Bunch-Davis vacuum as initial state. 

We studied very simple models of inflation that predict a featureless power spectrum at tree-level. 
Even for this case, the departure from the nearly scale-invariance appears at first-order in the perturbative 
expansion. For the $ \lambda \phi^4 $ model, the first-order radiative corrections to the primordial spectrum are 
proportional to the coupling of  the quartic interaction term, of the order of $10^{-13}$, so that the effect
is very small and we found a correction $\left|\frac{\Delta \mathcal{P}_R(k)}{\mathcal{P}^0_R(k)}\right|\leq 10^{-13}$. 
This is strongly dependent on the IR cut-off implemented and much larger values can be reached with other
IR cut-offs, i.e. the value of the scale factor at the beginning of inflation as used in~\cite{Sloth:2006az, Riotto:2008mv}.

The second type of models that we investigated is hybrid models, in particular those inspired by a supersymmetric theory. 
In this case the coupling can take much larger values up to order $ \lambda_h \sim 10^{-3}$ and the field in the loop is very 
massive, so that we avoid any IR divergence and dependence on the IR cut-off. We compute this contribution for
the first time using the full de Sitter propagator and computing the counter-terms analytically in the WKB approximation. 
Both the large coupling and mass conspire together in increasing the size of the one-loop correction and so we 
obtained a more optimistic value $\left|\frac{\Delta \mathcal{P}_{\cal R}(k)}{\mathcal{P}^0_{\cal R} (k)}
\right|\leq 10^{-1}$. 
Unfortunately, even in this example the oscillations are too small to be observed, or excluded by the present data.

Finally we investigated the loop effects on the trispectrum and even there the correction is too suppressed for
the chaotic case, but could be quite substantial for the hybrid models or for spectator fields, as long as
no cancellations occur. Note, though, that at least a partial cancellation is expected in the case of supersymmetric 
models, since we expect the cancellation of the UV  divergences in the loop in the limit of unbroken supersymmetry.

The latest Planck data can be well-fitted with a power-law power spectrum without any particular features~\cite{Ade:2015lrj}.
Nevertheless, a departure from scale invariance is still allowed due to the weak constraints on featured power spectra. 
In the future we expect progress in the reconstruction of the primordial universe and stronger constraints on the 
power spectra, giving the hope to maybe see the imprint of radiative corrections for sufficiently large coupling, as
in the case of hybrid inflation. In that case the shape of the oscillations could in principle be used to differentiate 
among other effects giving rise to oscillatory features in the spectrum.

\section*{Acknowledgements} 
The authors would like to thank Dorothea Bahns, Chiara Entradi and Thomas P. Hack for useful discussions.
LC would like to thank the Harish-Chandra Research Institute for hospitality during the final stages of this work.

This project has received funding from the German Research Foundation (Deutsche
Forschungsgemeinschaft (DFG)) through the Institutional Strategy of the
University of Göttingen (RTG 1493), the European Union's Horizon
2020 research and innovation programme InvisiblesPlus RISE under the 
Marie Sklodowska-Curie grant  agreement  No  690575, and from  the  
European Union's Horizon  2020  research  and  innovation programme  
Elusives  ITN  under  the Marie  Sklodowska-Curie grant agreement No 674896.

\begin{appendices}
\section{Feynman rules}\label{ap:FeynmanRules}
\subsection{\texorpdfstring{$\lambda \phi^4$}{lphi4}-theory}
The action for a $\frac{\lambda}{4!} \phi^4$-interacting field theory (\ref{eq:lagrang}) in the CTP formalism is given by
\begin{equation}
S[\phi^+, \phi^-] = \int_{\tau_\text{in}}^\infty \int d^3 x \,\mathcal{L}[\phi^+,\phi^-],
\end{equation}
where $\tau_\text{in}$ is the initial time where we assumed the system to be in a vacuum state, $\phi^+$ and $\phi^-$ are the 
two components of the field according to the CTP notation and 
\begin{multline}
\mathcal{L}[\phi^+,\phi^-] = \sqrt{-g}\Bigg[ \frac{1}{2} \partial_\mu \phi^+ \partial^\mu \phi^+ -\frac{1}{2}m^2(\phi^+)^2-\frac{\lambda}{4!} (\phi^+)^4 
+ \frac{\xi}{2} \;\mathcal{R} \; (\phi^+)^2  \\
- \frac{1}{2} \partial_\mu \phi^- \partial^\mu \phi^- +\frac{1}{2}m^2(\phi^-)^2+\frac{\lambda}{4!} (\phi^-)^4 
- \frac{\xi}{2} \;\mathcal{R} \; (\phi^-)^2 \Bigg]+\delta\mathcal{L}.
\end{multline}

\begin{table}[h!t]
\centering
\caption{Feynman rules for the $\lambda \phi^4$ theory. The scalar field $\phi^{(1)}$ is represented by a solid line and the field $\phi^{(2)}$ by a dotted line.}\label{tab:feynam}
\begin{tabular}{c|c}
\\
Graph&Expression\\
\hline \hline \raisebox{14mm}{}
   \input{fr-F.tex}&\raisebox{3mm}{$F(k,\tau_1,\tau_2)$}\\[2mm]
   \raisebox{14mm}{}
  	\input{fr-GR.tex}&\raisebox{3mm}{$-i G^R(k,\tau_1,\tau_2)=-i G^A(k,\tau_2,\tau_1)$}\\[2mm]
   \raisebox{14mm}{}
   \input{fr-coupling.tex}&\raisebox{3mm}{$-i a^4(\tau_1)\frac{\lambda}{4!}\delta(\tau_1-\tau_2)\delta(\tau_1-\tau_3)\delta(\tau_1-\tau_4)$}\\[2mm]
   \raisebox{14mm}{}
   	\input{fr-coupling1.tex}&\raisebox{3mm}{$-i a^4(\tau_1)\frac{\lambda}{3!}\delta(\tau_1-\tau_2)\delta(\tau_1-\tau_3)\delta(\tau_1-\tau_4)$}\\[2mm]
   \raisebox{14mm}{}
   \input{fr-F-count.tex}&\raisebox{3mm}{$-i a^4(\tau_1)\delta m^2 \delta(\tau_1-\tau_2)$}\\[2mm]
      \raisebox{14mm}{}
   \input{fr-coupling-count.tex}&\raisebox{3mm}{$-i a^4(\tau_1)\frac{\delta \lambda}{4!}\delta(\tau_1-\tau_2)\delta(\tau_1-\tau_3)\delta(\tau_1-\tau_4)$}\\[2mm]
         \raisebox{14mm}{}
   \input{fr-coupling1-count.tex}&\raisebox{3mm}{$-i a^4(\tau_1)\frac{\delta \lambda}{3!}\delta(\tau_1-\tau_2)\delta(\tau_1-\tau_3)\delta(\tau_1-\tau_4)$}\\[2mm]
\hline
\end{tabular}
\end{table}
In the previous expression the metric $g_{\mu\nu}$ has signature $+---$, $g=\det(g_{\mu\nu})$ and the counter-terms are defined as 
\begin{equation}
\delta\mathcal{L} =\sqrt{-g}\left( \frac{1}{2}\delta Z\; \partial_\mu \phi^+ \partial^\mu \phi^+ -\frac{1}{2}\delta m^2\; (\phi^+)^2-\frac{\delta\lambda}{4!} (\phi^+)^4 
+ \frac{\delta \xi}{2} \;\mathcal{R} \; (\phi^+)^2 
 \right) - \phi^+ \leftrightarrow \phi^-\,.
\end{equation} 
In the Schwinger basis (\ref{eq:change_of_basis}) the Lagrangian reads
\[
\mathcal{L}[\phi^{(1)},\phi^{(2)}] = \sqrt{-g}\left(\partial_\mu \phi^{(1)} \partial^\mu \phi^{(2)}  -m^2\phi^{(1)} \phi^{(2)} -\frac{\lambda}{4!} \phi^{(1)}(\phi^{(2)})^3 -\frac{\lambda}{3!} (\phi^{(1)})^3\phi^{(2)} + \frac{\xi}{3} \;\mathcal{R} \; \phi^{(1)} \phi^{(2)} \right)+ \delta\mathcal{L} .
\]
From the Lagrangian one can easily extract the CTP Feynman rules for the perturbative analysis. In Table~\ref{tab:feynam} 
we list the rules for the two-point functions, the self-interacting vertices and the counter-terms for $\xi =0$. There we 
represent $\phi^{(1)}$ with a solid line and $\phi^{(2)}$ with a dotted line.

\subsection{Hybrid model}
\begin{table}[h!t]
\centering
\caption{Feynman rules for the two scalar field theory (\ref{eq:CPThybridmod}). In the expressions we omitted the Dirac delta functions that can be reconstructed from Table \ref{tab:feynam}. The light fields $\phi^{(1)}$ and $\phi^{(2)}$ are represented by a solid and dotted single line, the heavy fields $\sigma^{(1)}$ and $\sigma^{(2)}$ by a solid and dotted double line.}\label{tab:feynamhyb}
\begin{tabular}{c|c|c|c}
\\
Graph&Expression&Graph&Expression\\
\hline \hline
\raisebox{14mm}{}
   \input{fr-F.tex}&\raisebox{3mm}{$F_\phi(k,\tau_1,\tau_2)$}&\input{fr-F_hyp.tex}&\raisebox{3mm}{$F_\sigma(k,\tau_1,\tau_2)$}\\[2mm]
\raisebox{14mm}{}
  	\input{fr-GR.tex}&\raisebox{3mm}{$-i G^R_\phi(k,\tau_1,\tau_2)$}&\input{fr-GR_hyb.tex}&\raisebox{3mm}{$-i G^R_\sigma(k,\tau_1,\tau_2)$}\\[2mm]
\raisebox{14mm}{}
   \input{fr-coupling_hyb.tex}&\raisebox{3mm}{$-i a^4(\tau_1)\frac{\lambda_h^2}{2}$}&\input{fr-coupling1_hyb.tex}&\raisebox{3mm}{$-2i a^4(\tau_1)\,\lambda_h^2\,$}\\[2mm]
\raisebox{14mm}{}
   \input{fr-coupling3_hyb.tex}&\raisebox{3mm}{$-i a^4(\tau_1)\frac{\lambda_h^2}{2}$}&\input{fr-coupling4_hyb.tex}&\raisebox{3mm}{$-2i a^4(\tau_1)\,\lambda_h^2\,$}\\[2mm]
   \raisebox{14mm}{}
   \input{fr-F-count.tex}&\raisebox{3mm}{$-i a^4(\tau_1)\delta m^2$}&\input{fr-F_hyb-count.tex}&\raisebox{3mm}{$-i a^4(\tau_1)\delta M^2$}\\[2mm]
\raisebox{14mm}{}
\raisebox{14mm}{}
   \input{fr-coupling_hyb-count.tex}&\raisebox{3mm}{$-i a^4(\tau_1)\frac{\delta \lambda_h^2}{2}$}&\input{fr-coupling1_hyb-count.tex}&\raisebox{3mm}{$-2i a^4(\tau_1)\,\delta \lambda_h^2$}\\[2mm]
\raisebox{14mm}{}
   \input{fr-coupling3_hyb-count.tex}&\raisebox{3mm}{$-i a^4(\tau_1)\frac{\delta \lambda_h^2}{2}$}&\input{fr-coupling4_hyb-count.tex}&\raisebox{3mm}{$-2i a^4(\tau_1)\,\delta \lambda_h^2$}\\[2mm]
\hline
\end{tabular}
\end{table}

We derive the Feynman rules for a 2-scalar field theory with interaction term $\lambda_h^2 \phi^2\sigma^2$. The \emph{in-in} action reads

\begin{equation}
S[\phi^+, \phi^-, \sigma^+,\sigma^-] = \int_{\tau_\text{in}}^\infty \int d^3 x \,\mathcal{L}[\phi^+, \phi^-, \sigma^+,\sigma^-],
\end{equation}
where we split also the second field in the two components $\sigma^+$ and $\sigma^-$. The Lagrangian is
\begin{multline}
\mathcal{L}[\phi^+, \phi^-, \sigma^+,\sigma^-] = \sqrt{-g}\Bigg[ \frac{1}{2} \partial_\mu \phi^+ \partial^\mu \phi^+ + \frac{1}{2} \partial_\mu \sigma^+ \partial^\mu \sigma^+ -\frac{1}{2}m^2(\phi^+)^2-\frac{1}{2}M^2(\sigma^+)^2-\lambda^2_h(\phi^+)^2 (\sigma^+)^2 \\
- \frac{1}{2} \partial_\mu \phi^- \partial^\mu \phi^- - \frac{1}{2} \partial_\mu \sigma^- \partial^\mu \sigma^- +\frac{1}{2}m^2(\phi^-)^2+\frac{1}{2}M^2(\sigma^-)^2+\lambda^2_h(\phi^-)^2 (\sigma^-)^2 \Bigg]+\delta\mathcal{L}.
\end{multline}

The counter-terms are defined as
\begin{multline}
\delta\mathcal{L} =\sqrt{-g}\Bigg( \frac{1}{2}\delta Z_\phi\; \partial_\mu \phi^+ \partial^\mu \phi^+ + \frac{1}{2}\delta Z_\sigma\; \partial_\mu \sigma^+ \partial^\mu \sigma^+\\
-\frac{1}{2}\delta m^2\; (\phi^+)^2  -\frac{1}{2}\delta M^2\; (\sigma^+)^2  -\delta\lambda^2_h (\phi^+)^2  (\sigma^+)^2  
 \Bigg) - \{\phi^+,\sigma^+\} \leftrightarrow \{\phi^-,\sigma^-\}\,.
\end{multline} 
In the Schwinger basis (\ref{eq:change_of_basis}) the Lagrangian reads
\begin{multline}\label{eq:CPThybridmod}
\mathcal{L}[\phi^{(1)},\phi^{(2)},\sigma^{(1)},\sigma^{(2)}] = \sqrt{-g}\Bigg(\partial_\mu \phi^{(1)} \partial^\mu \phi^{(2)}+\partial_\mu \sigma^{(1)} \partial^\mu \sigma^{(2)}  -m^2\phi^{(1)} \phi^{(2)}-M^2\sigma^{(1)} \sigma^{(2)}\\
 -2\lambda_h^2 \phi^{(1)}\phi^{(2)}(\sigma^{(1)})^2-2\lambda_h^2 (\phi^{(1)})^2\sigma^{(1)} \sigma^{(2)} -\frac{\lambda^2_h}{2} (\phi^{(2)})^2\sigma^{(1)}\sigma^{(2)}  -\frac{\lambda_h^2}{2} \phi^{(1)}\phi^{(2)}(\sigma^{(2)})^2 \Bigg)+ \delta\mathcal{L} .
\end{multline}
From the Lagrangian again we can identify the CTP Feynman rules. In Table~\ref{tab:feynamhyb} 
we list the rules for the two-point functions, the self-interacting vertices and the counter-terms. We represent $\phi^{(1)}$ with a solid single line, $\phi^{(2)}$ with a dotted single line, $\sigma^{(1)}$ with a solid double line and $\sigma^{(2)}$ with a dotted double line.

\section{One loop contribution to the 4-point function} \label{ap:4point}

We give in Table~\ref{tab:1loop} here below all the loop diagrams that contribute to the 4-point function. 
The integration in the momentum variables cannot be performed analytically without any simplification of the propagators. 
In the following calculations we will assume that the external momenta are super-Hubble, i.e. with wavelengths 
above the Hubble radius and consider the virtual particle to be massless, as done in  \cite{vanderMeulen:2007ah}.

Because in the Lagrangian there is no term proportional to $(\phi^{(1)})^2(\phi^{(2)})^2$, we expect that the sum of the diagrams $C_1 + C_2 + B$ is finite. The remaining diagrams contain instead new UV divergences that
are renormalized with a counter-term for the coupling term. 
Since we do not consider here massive propagators, all diagrams also contain IR divergences that are regulated by 
$ \varepsilon $ as for the case of the two-point function.

\begin{table}[h!t]
\caption{Inequivalent Feynman diagrams for the one-loop correction to the 4-point function. 
In the last column $\tau_1$ and $\tau_2$ denote the time of the left and right vertex and $ \tau $ is the external time.}\label{tab:1loop}
\begin{tabular}{c|c|c|c|c|c}
\\
Id &Coeff & Graph&Loop&Ext. contrib.&Time constraints\\
\hline \hline \raisebox{14mm}{}
   \raisebox{3mm}{\bf $A_1$}&\raisebox{3mm}{36}&\input{fishA1.tex}&\raisebox{3mm}{$-i G^R F$}&\raisebox{3mm}{$-i G^R F F F$}&\raisebox{3mm}{$\tau_2<\tau_1, \, \tau_1<\tau$}\\[2mm]
   \raisebox{14mm}{}
   \raisebox{3mm}{\bf $A_{11}$}&\raisebox{3mm}{36}&	\input{fishA11.tex}&\raisebox{3mm}{$-iG^A F$}&\raisebox{3mm}{$-i F F G^A F$}&\raisebox{3mm}{$\tau_1<\tau_2, \, \tau_2<\tau$}\\[2mm]
   \raisebox{14mm}{}
   \raisebox{3mm}{\bf $A_2$}&\raisebox{3mm}{36}& \input{fishA2.tex}&\raisebox{3mm}{$-i G^R F$}&\raisebox{3mm}{$iG^R FG^A G^A$}&\raisebox{3mm}{$\tau_2<\tau_1, \, \tau_1<\tau$}\\[2mm]
   \raisebox{14mm}{}
   \raisebox{3mm}{\bf $A_{21}$}&\raisebox{3mm}{36}&	\input{fishA21.tex}&\raisebox{3mm}{$-iG^A F$}&\raisebox{3mm}{$iG^R G^R G^A F$}&\raisebox{3mm}{$\tau_1<\tau_2, \, \tau_2<\tau$}\\[2mm]
   \raisebox{14mm}{}
   \raisebox{3mm}{\bf $B$ }&\raisebox{3mm}{18}&\input{fishB.tex} &\raisebox{3mm}{$F F$}&\raisebox{3mm}{$-G^R F G^A F$}&\raisebox{3mm}{$\tau_1<\tau, \, \tau_2<\tau$}\\[2mm]
   \raisebox{14mm}{}
   \raisebox{3mm}{\bf $C_1$}&\raisebox{3mm}{18}&\input{fishC.tex}&\raisebox{3mm}{$-G^R G^R$}&\raisebox{3mm}{$-G^R F G^A F$}&\raisebox{3mm}{$\tau_2<\tau_1, \, \tau_1<\tau$}\\[2mm]
   \raisebox{14mm}{}
   \raisebox{3mm}{\bf $C_2$}&\raisebox{3mm}{18}&\input{fishC2.tex} &\raisebox{3mm}{$-G^A G^A$}&\raisebox{3mm}{$-G^R F G^A F$}&\raisebox{3mm}{$\tau_1<\tau_2, \, \tau_2<\tau$}\\[2mm]
\hline
\end{tabular}
\end{table}

\subsection{Contribution \texorpdfstring{$B+C_1+C_2$}{BC1C2}}\label{ap:4pointBC}

Let us consider the finite contribution of the 4-point function of diagrams $B$, $C_1$ and $C_2$ in Table~\ref{tab:1loop}. 
In this case the product of distributions $F^2$, $(G^R)^2$ and $(G^A)^2$ diverge linearly, but the sum $(G^A)^2 + (G^R)^2 - F^2$ is finite.
Both are the same as the first-order correction to the two-point function for a $\lambda \phi^3$-scalar field theory and they were computed in \cite{vanderMeulen:2007ah}. We will use their results in the following.

To perform the momentum integration of the two propagators $F^2$ of diagram B one can split the integral in two parts: the small momentum contribution where the Hankel propagators are taken as in eqs.~(\ref{eq_dSp1},\ref{eq_dSp2}) with $ \nu = 3/2 - \varepsilon $ and the 
large momentum contribution. The small momentum contribution reads
\begin{equation}
\frac{H^4 \left(2 \log \left(k^2 \tau_1 \tau_2\right)+\frac{k}{M}-\frac{1}{2} \log \left(\frac{k+M}{M-k}\right)+\frac{1}{\varepsilon }\right)}{8 \pi ^2 k^3},
\end{equation}
where  $k = k_1+k_2$ and $M$ is the mass scale used as the upper limit for the momentum integral. The large momentum is
\begin{equation}
\frac{H^4 \left(\frac{1}{2} \log \left(\frac{k+M}{M-k}\right)-\frac{k}{M}\right)}{8 \pi ^2 k^3}-\frac{H^3 \Lambda  \tau_1^2 \tau_2^2 \sin (k (\tau_1-\tau_2))}{16 \pi ^2 k \tau_x (\tau_1-\tau_2)}.
\end{equation}
Here $\tau_x$ comes from the physical cutoff $-\Lambda/(H\tau_x)$, where $\tau_x=\min(\tau_1, \tau_2)$.  This contribution is linearly divergent in the cutoff $\Lambda$. Diagrams $C_1$ and $C_2$ will give similar expressions and the sum of the three diagrams will exactly cancel the linear divergence.

The amputated contribution $[B+C_1+C_2]_{\rm amp}$ is finally given by the sum of all small and large momenta contributions with the inclusion of two vertices $\lambda^2 a^4(\tau_1) a^4(\tau_2)$ with the correct combinatorial coefficients listed in Table \ref{tab:1loop}
\begin{equation}
-\frac{\lambda ^2 \left(2   \log \left((k_1+k_2)^2 \tau_1 \tau_2\right)+\frac{1}{\varepsilon}\right)}{288 \pi ^2 H^4 (k_1+k_2)^3 \tau_1^4 \tau_2^4  }.
\end{equation}
As expected the sum is finite in the ultraviolet regime and infrared divergent.

It is finally possible to consider the full diagrams with the contribution of the external propagators. To perform the time integrals analytically 
we assumed the super-Hubble approximation for the external momenta and obtained
\begin{multline}
\frac{H^4 \lambda ^2 \left(k_1^3+k_2^3\right) \left(k_3^3+k_4^3\right) \left(3 \log \left(\frac{\tau}{\tau_\text{in}}\right)+1\right)}{15552 \pi ^2 (k_1+k_2)^3 k_1^3 k_2^3 k_3^3 k_4^3  } \Big(9 \log \left(\frac{\tau}{\tau_\text{in}}\right) \left(2 \log \left(k^2 \tau \,\tau_\text{in}\right)+\frac{1}{\varepsilon}\right)\\
+12   \log (-k \tau)+4  +\frac{3}{\varepsilon}\Big).
\end{multline}

\subsection{Contribution \texorpdfstring{$A_i$}{Ai}} \label{ap:4pointA}

Similarly to what it has been done for diagrams $B$ and $C_i$~s, we study the contribution of diagrams $A_i$~s, that involve the product of the propagators $F$ and $G^{A/R}$. The UV behavior was already investigated in \cite{vanderMeulen:2007ah} for a $\lambda \phi^3$-theory. They found an analytic expression for the renormalized product $F \,G^R$, that in our setup corresponds to the renormalized amputated 4-point function
\begin{multline}
-\frac{i H^4}{24 \pi ^2} \theta (\tau_1-\tau_2) \Bigg[2\tau_1^3 \log \left(\frac{\left| \frac{\tau_2}{\tau_1-\tau_2}\right| }{2}\right)\\
+\frac{3}{2} \tau_1^2 \tau_2^2 \left(\delta (\tau_1-\tau_2) \left(\log \left(-\frac{2 \eta  \mu }{H \tau_1}\right)+\gamma \right)+\frac{\theta (-\eta +\tau_1-\tau_2)}{\tau_1-\tau_2}\right)\\
-2 \tau_2^3 \log \left(\frac{\tau_1}{2 (\tau_1-\tau_2)}\right)-2 \tau_1\tau_2 ( \tau_1- \tau_2)+\left(\frac{1}{\epsilon }+\frac{14}{3}-2 \gamma \right) \left( \tau_1^3- \tau_2^3\right)\Bigg],
\end{multline}
where again $\varepsilon$ is the mass regulator that has been introduced to cure the infrared divergence and $\mu$ is the renormalization scale. 
The ultraviolet divergence has been regulated by introducing the coupling constant counter-term
\begin{equation}
\delta \lambda = - 3\frac{\lambda^2}{16 \pi^2} \log{\frac{\Lambda}{\mu}},
\end{equation}
which fully agrees with the Minwkowski counter-term \cite{Ramond:1981pw} and the computations in different schemes as in
 \cite{Baacke:2010bm}, \cite{Markkanen:2013nwa}.

Now we will show our results for the fish diagram including the external propagators, that we assumed to be in a super-Hubble regime, i.e. $|k_i \tau|\ll1$. We found that the sum $A_1+A_{11}$ gives the contribution
\begin{multline}
-\frac{H^4 \lambda ^2 \left(k_1^3+k_2^3+k_3^3+k_4^3\right)}{62208 \pi ^2 k_1^3 k_2^3 k_3^3 k_4^3  } \Bigg(6 \log \left(\frac{\tau}{\tau_\text{in}}\right) \left(27   \log \left(\frac{\mu }{H}\right)+  6 \pi ^2+51 \gamma -127+51 \log (2)-\frac{12}{\varepsilon}\right) \\
+54   \left(\log \left(\frac{\mu}{H}\right)+4 \zeta (3)\right)+18 \log ^2\left(\frac{\tau}{\tau_\text{in}}\right) \left(2   \log \left(\frac{8 \tau_\text{in}}{\tau}\right)+2 (3 \gamma -8)  -\frac{3}{\varepsilon}\right)\\
+  51 \pi ^2+126 \gamma -776+126 \log (2)-\frac{36}{\varepsilon}\Bigg).
\end{multline}

Similarly, the sum of the two renormalized diagrams $A_2+A_{22}$ gives the contribution
\begin{multline}
-\frac{H^4 \lambda ^2 \tau_\text{in}^6 \left(k_1^3k_2^3k_3^3 + k_1^3k_2^3k_4^3 + k_1^3k_3^3k_4^3 +k_2^3 k_3^3 k_4^3\right)}{186624 \pi ^2 k_1^3 k_2^3 k_3^3 k_4^3 } \Bigg(9   \log \left(\frac{\mu }{H}\right)+\frac{1}{\varepsilon}\log \left(\frac{\tau_\text{in}^6}{\tau^6}\right)+\\
2   \left(3 \log \left(\frac{4 \tau_\text{in}}{\tau}\right)+6 \gamma -17\right) \log \left(\frac{\tau}{\tau_\text{in}}\right)+  2 \pi ^2+19 \gamma -48+19 \log (2)-\frac{5}{\varepsilon}\Bigg).
\end{multline}
As for the 2-point function we observe a logarithmic dependence on the conformal time and the first-order contributions of the series
expansion of the oscillatory terms.

\end{appendices}


\end{document}

%% file: CTP_profile.tikz
\draw[->,blue] (-2.15,0) -- (2.5,0);
\draw[->-,red,thick=1] (-1.5,0.2) -- (2.,0.2);
\draw[-<-=0.5,red,thick=1] (-1.5,-0.2) --  (2.,-0.2);
\draw[red,dashed,thick=1] (2,-0.2) arc (-90:90:0.2);

\draw[xshift=-1.5 cm] (0pt,1pt) -- (0pt,-1pt) node[left,yshift=-2.1,scale=0.8] {$t_{in}$};
\draw[xshift=0.5 cm] (0pt,1pt) -- (0pt,-1pt) node[left,yshift=-2.1,scale=0.8] {$t$};

\node[red, text width=3cm,yshift=4,scale=0.8] at (-1.7,0) {TIME};
\node[blue, text width=3cm,yshift=4.2,scale=0.8] at (3,0) {$+\infty$};
\node[blue, text width=3cm,yshift=1.,scale=0.8] at (-0.5,0.3) {$\phi^+$};
\node[blue, text width=3cm,yshift=-1.,scale=0.8] at (2.2,-0.3) {$\phi^-$};

%% file: propag++.tex
\begin{fmffile}{propagatore++}
\begin{fmfgraph*}(50,40)
\fmfleft{i}
\fmfright{o}
\fmf{plain}{i,i1}
\fmf{plain}{i1,o}
\end{fmfgraph*}
\end{fmffile}

%% file: tadpoleA.tex
\begin{fmffile}{tadpoleA}
\begin{fmfgraph*}(70,30)
\fmfleft{i}
\fmfright{o}
\fmflabel{$k_1$}{i}
\fmflabel{$k_2$}{o}
\fmf{plain}{v,v}
\fmf{plain}{v,o}
\fmf{phantom,tag=1}{i,v}
\fmfv{label=$\tau_1$,label.dist=1.2mm, label.angle=-90}{v}
\fmfposition
\fmfipath{p[]}
\fmfiset{p1}{vpath1(__i,__v)}
\fmfi{plain}{subpath (0,length(p1)/2) of p1}
\fmfi{dots}{subpath (length(p1)/2,length(p1)) of p1}
\end{fmfgraph*}
\end{fmffile}

%% file: tadpoleAc.tex
\begin{fmffile}{tadpoleAc}
\begin{fmfgraph*}(70,30)
\fmfleft{i}
\fmfright{o}
\fmflabel{$k_1$}{i}
\fmflabel{$k_2$}{o}
\fmfv{decor.shape=cross,decor.size=9thick}{v}
\fmf{plain}{v,o}
\fmf{phantom,tag=1}{i,v}
\fmfposition
\fmfipath{p[]}
\fmfiset{p1}{vpath1(__i,__v)}
\fmfi{plain}{subpath (0,length(p1)/2) of p1}
\fmfi{dots}{subpath (length(p1)/2,length(p1)) of p1}
\end{fmfgraph*}
\end{fmffile}

%% file: tree1.tex
\begin{fmffile}{tree1}
\begin{fmfgraph*}(50,40)
\fmfleft{i1,i2}
\fmfright{o1,o2}
\fmfv{label=$\tau_1$,label.dist=1.9mm, label.angle=0}{v}
\fmf{plain}{i1,i1i}
\fmf{plain}{i1i,v}
\fmf{plain}{i2,i2i}
\fmf{dots}{i2i,v}
\fmf{plain}{v,o1i}
\fmf{plain}{o1i,o1}
\fmf{plain}{v,o2i}
\fmf{plain}{o2i,o2}
\end{fmfgraph*}
\end{fmffile}

%% file: tree2.tex
\begin{fmffile}{tree2}
\begin{fmfgraph}(50,40)
\fmfleft{i1,i2}
\fmfright{o1,o2}
\fmf{plain}{i1,i1i}
\fmf{plain}{i1i,v}
\fmf{plain}{i2,i2i}
\fmf{dots}{i2i,v}
\fmf{dots}{v,o1i}
\fmf{plain}{o1i,o1}
\fmf{dots}{v,o2i}
\fmf{plain}{o2i,o2}
\end{fmfgraph}
\end{fmffile}

%% file: fr-F.tex
\begin{fmffile}{fr-F}
\begin{fmfgraph*}(70,30)
\fmfleft{i}
\fmfright{o}
\fmf{plain}{i,v}
\fmf{plain}{v,o}
\end{fmfgraph*}
\end{fmffile}

%% file: fr-GR.tex
\begin{fmffile}{fr-GR}
\begin{fmfgraph*}(70,30)
\fmfleft{i}
\fmfright{o}
\fmf{plain}{i,v}
\fmf{dots}{v,o}
\end{fmfgraph*}
\end{fmffile}

%% file: fr-coupling.tex
\begin{fmffile}{fr-coupling}
\begin{fmfgraph}(50,40)

\fmfleft{i1,i2}
\fmfright{o1,o2}
\fmf{plain}{i1,v}
\fmf{dots}{i2,v}
\fmf{dots}{v,o1}
\fmf{dots}{v,o2}

\fmflabel{\(g\)}{i1}                    

\end{fmfgraph}
\end{fmffile}

%% file: fr-coupling1.tex
\begin{fmffile}{fr-coupling_sec}
\begin{fmfgraph*}(50,40)
\fmfleft{i1,i2}
\fmfright{o1,o2}
\fmfv{label=$\tau_1$,label.dist=1.9mm, label.angle=0}{v}
\fmf{plain}{i1,v}
\fmf{dots}{i2,v}
\fmf{plain}{v,o1}
\fmf{plain}{v,o2}
\end{fmfgraph*}
\end{fmffile}

%% file: fr-F-count.tex
\begin{fmffile}{fr-F-count}
\begin{fmfgraph*}(70,30)
\fmfleft{i}
\fmfright{o}
\fmfv{decor.shape=cross,decor.size=9thick}{v}
\fmf{dots}{v,o}
\fmf{plain}{i,v}

\end{fmfgraph*}
\end{fmffile}

%% file: fr-coupling-count.tex
\begin{fmffile}{fr-coupling-count}
\begin{fmfgraph}(50,40)
\fmfleft{i1,i2}
\fmfright{o1,o2}
\fmf{plain}{i1,v}
\fmfv{decor.shape=circle,decor.filled=empty,decor.size=3thick}{v}
\fmf{dots}{i2,v}
\fmf{dots}{v,o1}
\fmf{dots}{v,o2}
\end{fmfgraph}
\end{fmffile}

%% file: fr-coupling1-count.tex
\begin{fmffile}{fr-coupling_count_sec}
\begin{fmfgraph*}(50,40)
\fmfleft{i1,i2}
\fmfright{o1,o2}
\fmfv{label=$\tau_1$,label.dist=1.9mm, label.angle=0}{v}
\fmf{plain}{i1,v}
\fmfv{decor.shape=circle,decor.filled=empty,decor.size=3thick}{v}
\fmf{dots}{i2,v}
\fmf{plain}{v,o1}
\fmf{plain}{v,o2}
\end{fmfgraph*}
\end{fmffile}

%% file: fr-F_hyp.tex
\begin{fmffile}{fr-F_hyb}
\begin{fmfgraph*}(70,30)
\fmfleft{i}
\fmfright{o}
\fmf{dbl_plain}{i,v}
\fmf{dbl_plain}{v,o}
\end{fmfgraph*}
\end{fmffile}

%% file: fr-GR_hyb.tex
\begin{fmffile}{fr-GR_hyb}
\begin{fmfgraph*}(70,30)
\fmfleft{i}
\fmfright{o}
\fmf{dbl_plain}{i,v}
\fmf{dbl_dots}{v,o}
\end{fmfgraph*}
\end{fmffile}

%% file: fr-coupling_hyb.tex
\begin{fmffile}{fr-coupling_hyb}
\begin{fmfgraph}(50,40)

\fmfleft{i1,i2}
\fmfright{o1,o2}
\fmf{plain}{i1,v}
\fmf{dots}{i2,v}
\fmf{dbl_dots}{v,o1}
\fmf{dbl_dots}{v,o2}

\fmflabel{\(g\)}{i1}                    

\end{fmfgraph}
\end{fmffile}

%% file: fr-coupling1_hyb.tex
\begin{fmffile}{fr-coupling1_hyb}
\begin{fmfgraph*}(50,40)
\fmfleft{i1,i2}
\fmfright{o1,o2}
\fmfv{label=$\tau_1$,label.dist=1.9mm, label.angle=0}{v}
\fmf{plain}{i1,v}
\fmf{dots}{i2,v}
\fmf{dbl_plain}{v,o1}
\fmf{dbl_plain}{v,o2}
\end{fmfgraph*}
\end{fmffile}

%% file: fr-coupling3_hyb.tex
\begin{fmffile}{fr-coupling3_hyb}
\begin{fmfgraph}(50,40)

\fmfleft{i1,i2}
\fmfright{o1,o2}
\fmf{dots}{i1,v}
\fmf{dots}{i2,v}
\fmf{dbl_dots}{v,o1}
\fmf{dbl_plain}{v,o2}

\fmflabel{\(g\)}{i1}                    

\end{fmfgraph}
\end{fmffile}

%% file: fr-coupling4_hyb.tex
\begin{fmffile}{fr-coupling4_hyb}
\begin{fmfgraph*}(50,40)
\fmfleft{i1,i2}
\fmfright{o1,o2}
\fmfv{label=$\tau_1$,label.dist=1.9mm, label.angle=0}{v}
\fmf{plain}{i1,v}
\fmf{plain}{i2,v}
\fmf{dbl_plain}{v,o1}
\fmf{dbl_dots}{v,o2}
\end{fmfgraph*}
\end{fmffile}

%% file: fr-F_hyb-count.tex
\begin{fmffile}{fr-F_hyb-count}
\begin{fmfgraph*}(70,30)
\fmfleft{i}
\fmfright{o}
\fmfv{decor.shape=cross,decor.size=9thick}{v}
\fmf{dbl_dots}{v,o}
\fmf{dbl_plain}{i,v}

\end{fmfgraph*}
\end{fmffile}

%% file: fr-coupling_hyb-count.tex
\begin{fmffile}{fr-coupling_hyb-count}
\begin{fmfgraph}(50,40)
\fmfleft{i1,i2}
\fmfright{o1,o2}
\fmf{plain}{i1,v}
\fmfv{decor.shape=circle,decor.filled=empty,decor.size=3thick}{v}
\fmf{dots}{i2,v}
\fmf{dbl_dots}{v,o1}
\fmf{dbl_dots}{v,o2}
\end{fmfgraph}
\end{fmffile}

%% file: fr-coupling1_hyb-count.tex
\begin{fmffile}{fr-coupling1_hyb_count_sec}
\begin{fmfgraph*}(50,40)
\fmfleft{i1,i2}
\fmfright{o1,o2}
\fmfv{label=$\tau_1$,label.dist=1.9mm, label.angle=0}{v}
\fmf{plain}{i1,v}
\fmfv{decor.shape=circle,decor.filled=empty,decor.size=3thick}{v}
\fmf{dots}{i2,v}
\fmf{dbl_plain}{v,o1}
\fmf{dbl_plain}{v,o2}
\end{fmfgraph*}
\end{fmffile}

%% file: fr-coupling3_hyb-count.tex
\begin{fmffile}{fr-coupling3_hyb-count}
\begin{fmfgraph}(50,40)
\fmfleft{i1,i2}
\fmfright{o1,o2}
\fmf{dots}{i1,v}
\fmfv{decor.shape=circle,decor.filled=empty,decor.size=3thick}{v}
\fmf{dots}{i2,v}
\fmf{dbl_dots}{v,o1}
\fmf{dbl_plain}{v,o2}
\end{fmfgraph}
\end{fmffile}

%% file: fr-coupling4_hyb-count.tex
\begin{fmffile}{fr-coupling4_hyb_count_sec}
\begin{fmfgraph*}(50,40)
\fmfleft{i1,i2}
\fmfright{o1,o2}
\fmfv{label=$\tau_1$,label.dist=1.9mm, label.angle=0}{v}
\fmf{plain}{i1,v}
\fmfv{decor.shape=circle,decor.filled=empty,decor.size=3thick}{v}
\fmf{plain}{i2,v}
\fmf{dbl_plain}{v,o1}
\fmf{dbl_dots}{v,o2}
\end{fmfgraph*}
\end{fmffile}

%% file: fishA1.tex
\begin{fmffile}{fishA1}
\begin{fmfgraph*}(120,30)
\fmfleft{i1,i2}
\fmfright{o1,o2}
\fmfv{label=$\tau_1$,label.dist=1.2mm, label.angle=0}{v1}
\fmfv{label=$\tau_2$,label.dist=1.2mm,label.angle=180}{v2}
\fmflabel{$k_2$}{i1}
\fmflabel{$k_1$}{i2}
\fmflabel{$k_4$}{o1}
\fmflabel{$k_3$}{o2}
\fmf{plain}{i1,i1i}
\fmf{plain}{i1i,v1}
\fmf{plain}{i2,i2i}
\fmf{dots}{i2i,v1}
\fmf{plain}{v2,o1i}
\fmf{plain}{o1i,o1}
\fmf{plain}{v2,o2i}
\fmf{plain}{o2i,o2}
\fmf{phantom,left=0.8,tension=0.4,tag=1}{v1,v2}
\fmf{phantom,left=0.8,tension=0.4,tag=2}{v2,v1}
\fmfposition
\fmfipath{p[]}
\fmfiset{p1}{vpath1(__v1,__v2)}
\fmfiset{p2}{vpath2(__v2,__v1)}
\fmfi{plain}{subpath (0,length(p1)/2) of p1}
\fmfi{dots}{subpath (length(p1)/2,length(p1)) of p1}
\fmfi{plain}{subpath (0,length(p2)/2) of p2}
\fmfi{plain}{subpath (length(p2)/2,length(p2)) of p2}
\end{fmfgraph*}
\end{fmffile}

%% file: fishA11.tex
\begin{fmffile}{fishA11}
\begin{fmfgraph}(120,30)
\fmfleft{i1,i2}
\fmfright{o1,o2}
\fmf{plain}{i1,i1i}
\fmf{plain}{i1i,v1}
\fmf{plain}{i2,i2i}
\fmf{plain}{i2i,v1}
\fmf{plain}{v2,o1i}
\fmf{plain}{o1i,o1}
\fmf{dots}{v2,o2i}
\fmf{plain}{o2i,o2}
\fmf{phantom,left=0.8,tension=0.4,tag=1}{v1,v2}
\fmf{phantom,left=0.8,tension=0.4,tag=2}{v2,v1}
\fmfposition
\fmfipath{p[]}
\fmfiset{p1}{vpath1(__v1,__v2)}
\fmfiset{p2}{vpath2(__v2,__v1)}
\fmfi{dots}{subpath (0,length(p1)/2) of p1}
\fmfi{plain}{subpath (length(p1)/2,length(p1)) of p1}
\fmfi{plain}{subpath (0,length(p2)/2) of p2}
\fmfi{plain}{subpath (length(p2)/2,length(p2)) of p2}
\end{fmfgraph}
\end{fmffile}

%% file: fishA2.tex
\begin{fmffile}{fishA2}
\begin{fmfgraph}(120,30)
\fmfleft{i1,i2}
\fmfright{o1,o2}
\fmf{plain}{i1,i1i}
\fmf{plain}{i1i,v1}
\fmf{plain}{i2,i2i}
\fmf{dots}{i2i,v1}
\fmf{dots}{v2,o1i}
\fmf{plain}{o1i,o1}
\fmf{dots}{v2,o2i}
\fmf{plain}{o2i,o2}
\fmf{phantom,left=0.8,tension=0.4,tag=1}{v1,v2}
\fmf{phantom,left=0.8,tension=0.4,tag=2}{v2,v1}
\fmfposition
\fmfipath{p[]}
\fmfiset{p1}{vpath1(__v1,__v2)}
\fmfiset{p2}{vpath2(__v2,__v1)}
\fmfi{plain}{subpath (0,length(p1)/2) of p1}
\fmfi{dots}{subpath (length(p1)/2,length(p1)) of p1}
\fmfi{plain}{subpath (0,length(p2)/2) of p2}
\fmfi{plain}{subpath (length(p2)/2,length(p2)) of p2}
\end{fmfgraph}
\end{fmffile}

%% file: fishA21.tex
\begin{fmffile}{fishA21}
\begin{fmfgraph}(120,30)
\fmfleft{i1,i2}
\fmfright{o1,o2}
\fmf{plain}{i1,i1i}
\fmf{dots}{i1i,v1}
\fmf{plain}{i2,i2i}
\fmf{dots}{i2i,v1}
\fmf{plain}{v2,o1i}
\fmf{plain}{o1i,o1}
\fmf{dots}{v2,o2i}
\fmf{plain}{o2i,o2}
\fmf{phantom,left=0.8,tension=0.4,tag=1}{v1,v2}
\fmf{phantom,left=0.8,tension=0.4,tag=2}{v2,v1}
\fmfposition
\fmfipath{p[]}
\fmfiset{p1}{vpath1(__v1,__v2)}
\fmfiset{p2}{vpath2(__v2,__v1)}
\fmfi{dots}{subpath (0,length(p1)/2) of p1}
\fmfi{plain}{subpath (length(p1)/2,length(p1)) of p1}
\fmfi{plain}{subpath (0,length(p2)/2) of p2}
\fmfi{plain}{subpath (length(p2)/2,length(p2)) of p2}
\end{fmfgraph}
\end{fmffile}

%% file: fishB.tex
\begin{fmffile}{fishB}
\begin{fmfgraph}(120,30)
\fmfleft{i1,i2}
\fmfright{o1,o2}
\fmf{plain}{i1,i1i}
\fmf{plain}{i1i,v1}
\fmf{plain}{i2,i2i}
\fmf{dots}{i2i,v1}
\fmf{plain}{v2,o1i}
\fmf{plain}{o1i,o1}
\fmf{dots}{v2,o2i}
\fmf{plain}{o2i,o2}
\fmf{phantom,left=0.8,tension=0.4,tag=1}{v1,v2}
\fmf{phantom,left=0.8,tension=0.4,tag=2}{v2,v1}
\fmfposition
\fmfipath{p[]}
\fmfiset{p1}{vpath1(__v1,__v2)}
\fmfiset{p2}{vpath2(__v2,__v1)}
\fmfi{plain}{subpath (0,length(p1)/2) of p1}
\fmfi{plain}{subpath (length(p1)/2,length(p1)) of p1}
\fmfi{plain}{subpath (0,length(p2)/2) of p2}
\fmfi{plain}{subpath (length(p2)/2,length(p2)) of p2}
\end{fmfgraph}
\end{fmffile}

%% file: fishC.tex
\begin{fmffile}{fishC1}
\begin{fmfgraph}(120,30)
\fmfleft{i1,i2}
\fmfright{o1,o2}
\fmf{plain}{i1,i1i}
\fmf{plain}{i1i,v1}
\fmf{plain}{i2,i2i}
\fmf{dots}{i2i,v1}
\fmf{plain}{v2,o1i}
\fmf{plain}{o1i,o1}
\fmf{dots}{v2,o2i}
\fmf{plain}{o2i,o2}
\fmf{phantom,left=0.8,tension=0.4,tag=1}{v1,v2}
\fmf{phantom,left=0.8,tension=0.4,tag=2}{v2,v1}
\fmfposition
\fmfipath{p[]}
\fmfiset{p1}{vpath1(__v1,__v2)}
\fmfiset{p2}{vpath2(__v2,__v1)}
\fmfi{plain}{subpath (0,length(p1)/2) of p1}
\fmfi{dots}{subpath (length(p1)/2,length(p1)) of p1}
\fmfi{dots}{subpath (0,length(p2)/2) of p2}
\fmfi{plain}{subpath (length(p2)/2,length(p2)) of p2}
\end{fmfgraph}
\end{fmffile}

%% file: fishC2.tex
\begin{fmffile}{fishC2}
\begin{fmfgraph}(120,30)
\fmfleft{i1,i2}
\fmfright{o1,o2}
\fmf{plain}{i1,i1i}
\fmf{plain}{i1i,v1}
\fmf{plain}{i2,i2i}
\fmf{dots}{i2i,v1}
\fmf{plain}{v2,o1i}
\fmf{plain}{o1i,o1}
\fmf{dots}{v2,o2i}
\fmf{plain}{o2i,o2}
\fmf{phantom,left=0.8,tension=0.4,tag=1}{v1,v2}
\fmf{phantom,left=0.8,tension=0.4,tag=2}{v2,v1}
\fmfposition
\fmfipath{p[]}
\fmfiset{p1}{vpath1(__v1,__v2)}
\fmfiset{p2}{vpath2(__v2,__v1)}
\fmfi{dots}{subpath (0,length(p1)/2) of p1}
\fmfi{plain}{subpath (length(p1)/2,length(p1)) of p1}
\fmfi{plain}{subpath (0,length(p2)/2) of p2}
\fmfi{dots}{subpath (length(p2)/2,length(p2)) of p2}
\end{fmfgraph}
\end{fmffile}